\documentclass[fleqn,10pt]{wlscirep}
\usepackage[utf8]{inputenc}
\usepackage[T1]{fontenc}

\usepackage{subfigure}
\usepackage{graphicx}
\usepackage{placeins} 
\usepackage{tikz}

\definecolor{mygreen}{RGB}{0 150 130} 
\definecolor{myred}{RGB}{162 34 35} 
\definecolor{myblue}{RGB}{70 100 170} 
\definecolor{myyellow}{RGB}{252 229 0} 
\definecolor{mygreen2}{RGB}{75 173 59} 
\definecolor{myred2}{RGB}{219 51 51}
\definecolor{myblack}{RGB}{0 0 0}

\title{Interferometric Particle Imaging for  Particle Sizing in the Front-, Side-, and Back-Scatter Region}

\author[1,*]{Christian Sax}
\author[1]{Maximilian Dreisbach}
\author[1]{Jochen Kriegseis}
\affil[1]{Karlsruhe Institute of Technology (KIT), Institute of Fluid Dynamics, Karlsruhe, D-76131, Germany}

\affil[*]{christian.sax@kit.edu}

\begin{abstract}
Interferometric particle imaging is a widely used optical measuring technique for the sizing of poly-dispersed spherical particles like droplets and bubbles. In its conventional approach, the method is limited to forward-scattering angles and therefore, requiring a second optical access, restricting the range of possible applications. In the present work, this limitation of the scattering angle is addressed, showing that also other scattering angles, especially in the back-scatter region are applicable, expanding the technique to applications with only a single optical access. A general method for the identification of suitable scattering angles both for droplets and bubbles is proposed. The visibility criterion for interference patterns from particles is generalized and possible glare point parings and their separation in the forward-, side- and back-scatter regimes are discussed for droplets and bubbles. Due to being the most popular examples, different scattering angles are proposed for water droplets and air bubbles in water. In the last part, the method is validated on a bubble sizing experiment.
\end{abstract}
\begin{document}

\flushbottom
\maketitle
%
%
\thispagestyle{empty}


\section*{Introduction}

Sizing of transparent spherical particles is of great interest in a wide variety of research fields. Such applications are the characterization of droplets e.g in sprays \cite{Maeda2002,Hardalupas2009} at the outlet of nozzles \cite{Matsuura2006}, in combustion \cite{Fujisawa2003} and droplets in clouds and atmosphere \cite{Dunker2016,KIELAR2016,Querel}. Likewise, the dynamic of gas bubbles in multi-phase flows are widely investigated e.g. cavitation and bubbly flow \cite{Lacagnina2011,Kawaguchi.2002,Niwa.2000}. Especially for small transparent particles, interferometric particle imaging (IPI) \cite{Konig.1986,Niwa.2000,Glover.95,Maeda2002} is a popular volumetric non-intrusive laser-optical measurement technique for the time resolved sizing of poly-dispersed particles. The technique utilizes the glare points from scattered laser light, visible on surface of the transparent particle, to determine the particle diameter. The light exiting the particle through the glare points, form an interference pattern when the particle is imaged out of focus. Consequently, the particle diameter can be determined from the interference pattern. IPI bears the advantage of only requiring a single camera for imaging and a single laser for illumination. Additionally the method is easily combined with particle tracking velocimetry (PTV) approaches without changes in the setup being necessary, allowing for simultaneous particle sizing and tracking \cite{Lacagnina2011}. In combination with defocusing particle tracking velocimetry (DPTV) \cite{Willert.1992,Fuchs.2016} even three dimensional tracking is possible with the single camera setup. IPI allows for the sizing of small particles, for which shadowgraphy based techniques can not obtain sufficient accuracy. Being a volumetric method IPI allows for a larger measurement volumes compared to single-point techniques like PDA\cite{Tropea2007}.\\
First introduced as interferometric laser imaging for droplet sizing (ILIDS) \cite{Konig.1986,Glover.95,Rousselle1999}, the method was developed for the sizing of droplets (particles with larger refractive index than the surrounding medium). Later on the technique was expanded to bubbles (particles with smaller refractive index than the surrounding medium)\cite{Niwa.2000,Kawaguchi.2002}. IPI is usually performed in the front scatter region, with an scattering angle of 66° for droplets and 45° for bubbles respectively. With the conventional IPI approach, using the reflected ($p=0$) and first order refracted ($p=1$) glare point, the method is limited to scattering angles being smaller than $2arccos(m)$, with $m$ being the relative reflective index, and therefore to the front-scatter region. This results in IPI having the drawback of requiring a second optical access, limiting this useful technique in its range of possible applications. More recently IPI was also performed in the side scatter regime at 90° by utilizing the second order ($p=2$) refracted glare point instead of the first order glare point \cite{Zhang.2018,Russell2020}. While this allows for more variation in the necessary scattering angle, the side scatter regime still requires a second optical access. With the utilization of the different glare point parings ($p=0\,\&\,p=2$ instead of $p=0\,\&\,p=1$) leading to more possible scattering angles, the question arises which other scattering angles are suitable for IPI and which glare points are the dominant ones for these angles. Especially the identification of applicable scattering angles for IPI in the back-scatter regime is highly desirable since it would remove the necessity of a second optical access, making IPI available for a wider variety of applications.

\section*{Interferometric Particle Imaging}
\label{Sec:IPI}

\begin{figure} [h]
\centering
 \addtocounter{subfigure}{0}
    \subfigure[Sphere $m\geq 1$ (droplet)]{\includegraphics[width=0.25\textwidth]{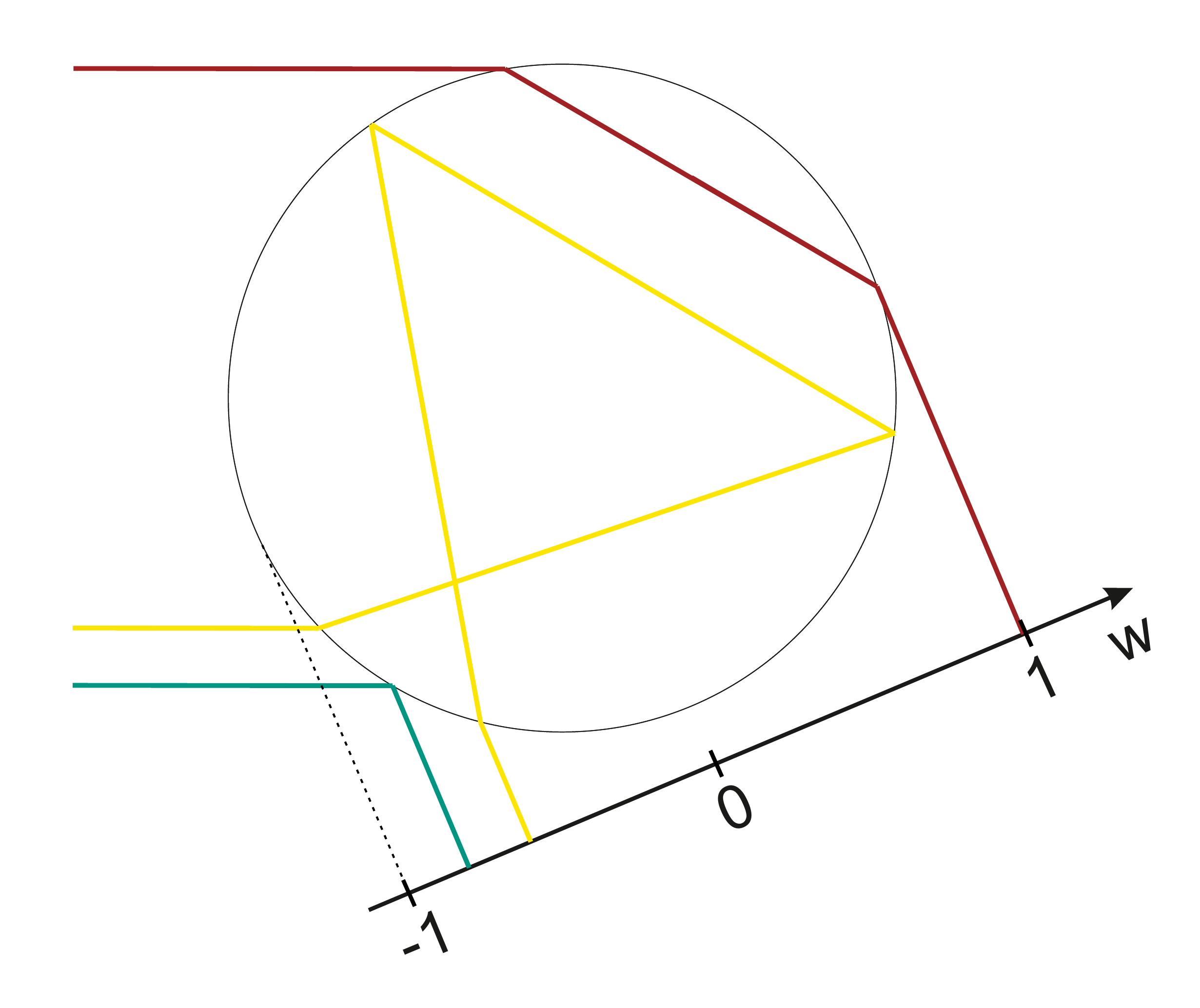}}
    \hspace{1cm}
    \subfigure[Sphere $m\leq 1$ (bubble)]{\includegraphics[width=0.25\textwidth]{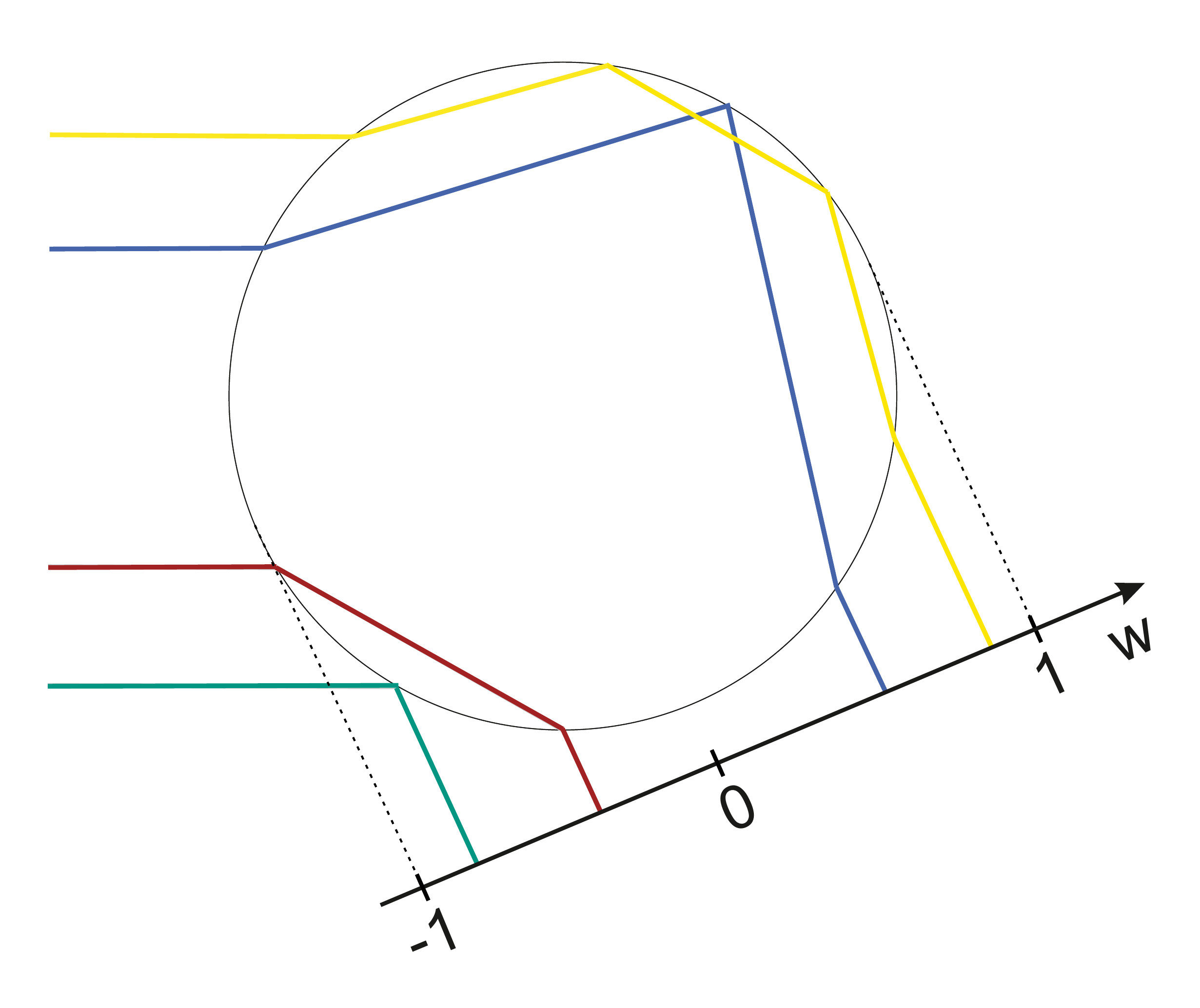}}
\caption{Geometrical optics approximation of light rays through a sphere with a relative reflective index a) $m>1$ (droplet) and b) $m<1$ (bubble). Depicted are the reflected ray $p=0$ ($\textcolor{mygreen}{\bullet}$) and the first three orders refracted rays $p=1$ ($\textcolor{myred}{\bullet}$), $p=2$ ($\textcolor{myblue}{\bullet}$) and $p=3$ ($\textcolor{myyellow}{\bullet}$). The respective glare point position on the spheres surface is shown by the $w$ coordinate. Note that the light is monochromatic and the colors encode the scattering order. Depicted is also the particle surface coordinate $w$ which describes the projected (visible) area of the particle from the observation angle $\theta$.}
\label{fig:IPI_GO_Sphere}
\end{figure}

\begin{figure} [h]
\centering
    \subfigure[Defocusing glare points]{\includegraphics[width=0.5\textwidth]{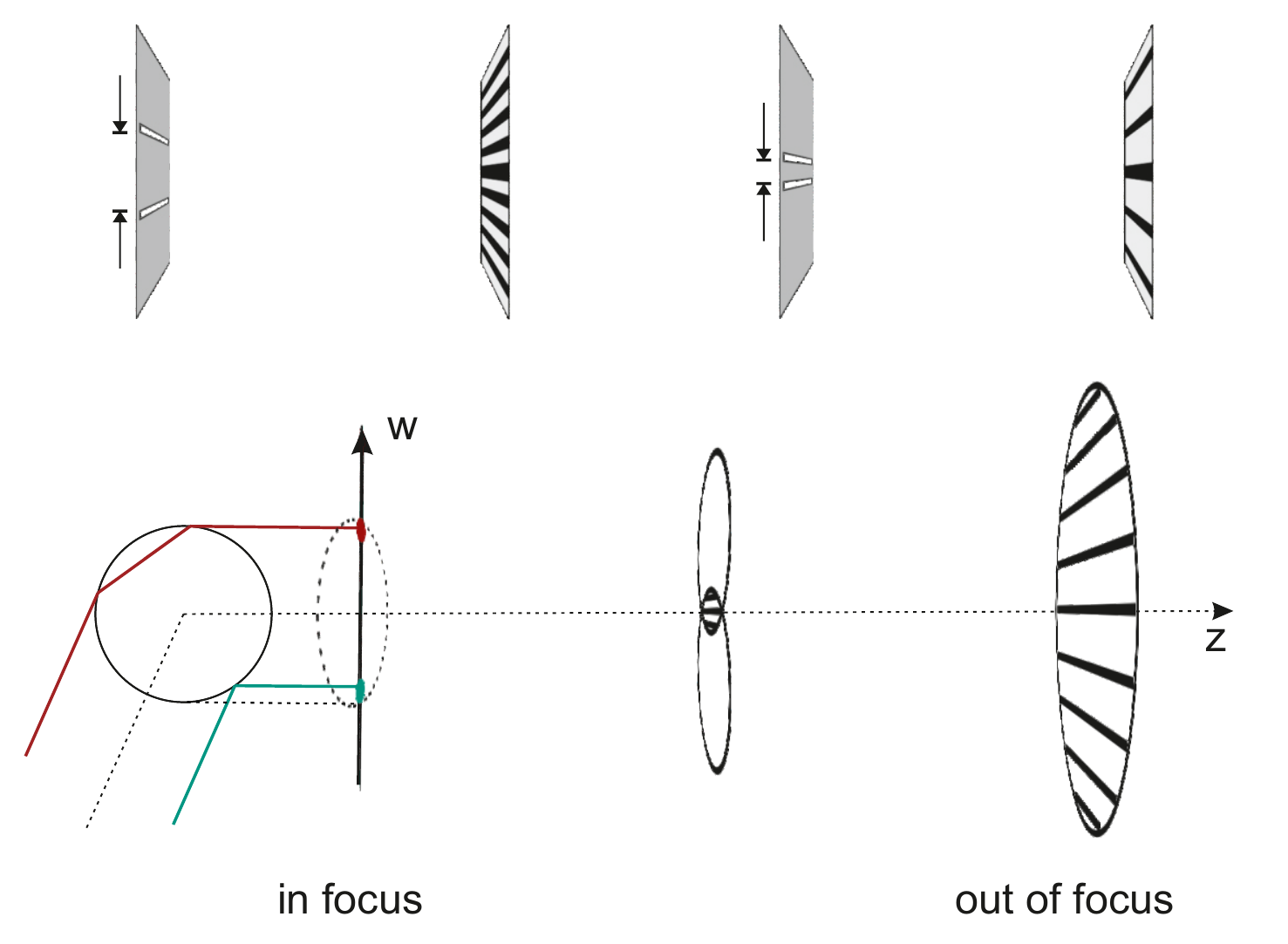}}
    \hspace{1cm}
    \subfigure[Angles with GO]{\hspace{1mm}\includegraphics[width=0.35\textwidth]{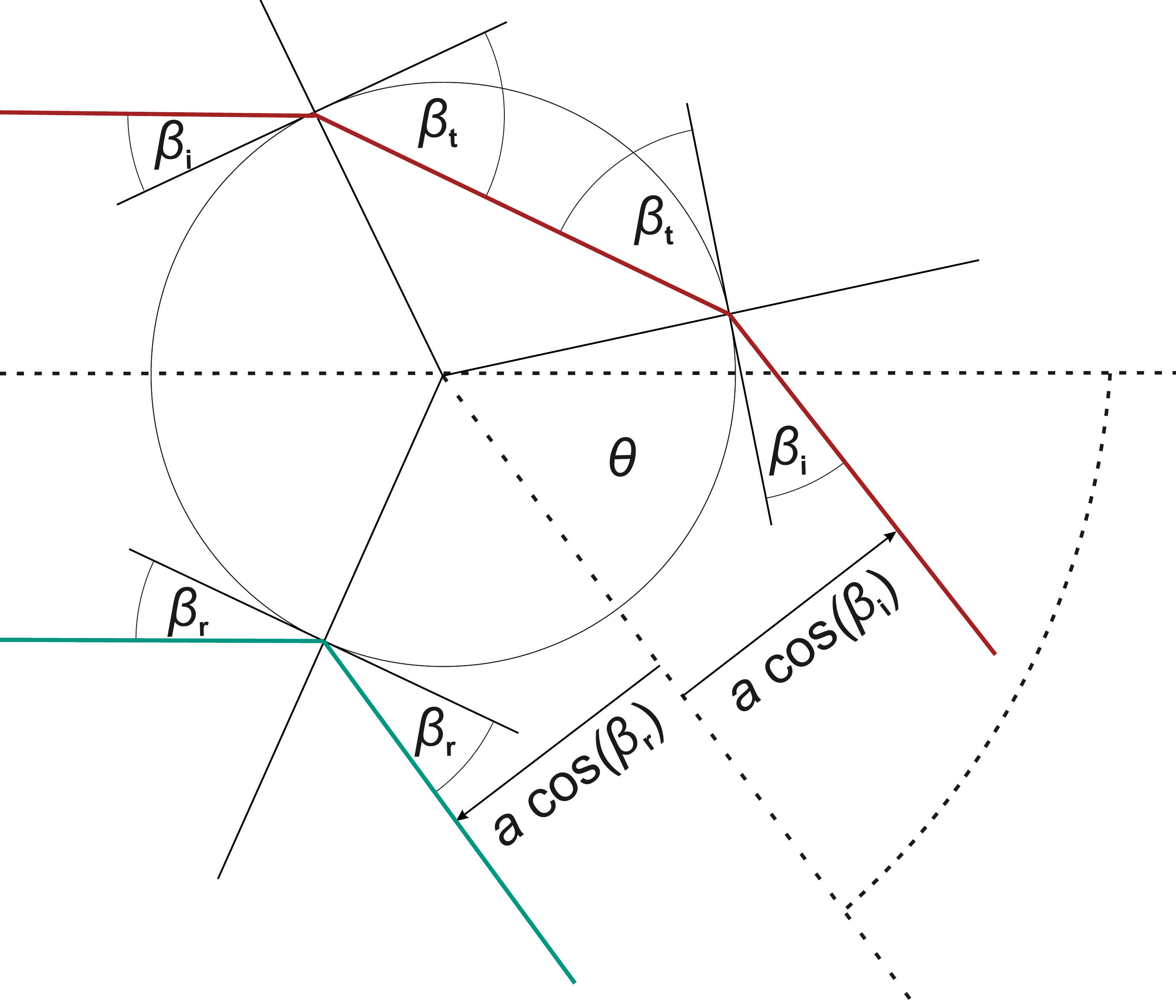}}
\caption{a) Working principle of IPI visualized. Two glare points are visible on the projected image of the bubble in focus from the observation angle theta. The image of the glare points start to overlap and form an interference pattern when moving away from the focal plane, until eventually collapsing into a singe circle. Sketch based on \cite{Madsen.2003}. Additionally the analogy of the glare point spacing (particle size) in IPI to a double slit in a Young's fringe experiment is visualized. The intensity maxima are depicted as dark stripes for better visibility. b) Tracing the light path with GO approximation for the determination of the glare point position, on the example of an droplet.}
\label{fig:IPI_WorkingPrinciple}
\end{figure}

Considering a transparent spherical particle illuminated with coherent monochromatic light, the light is scattered at the particle, causing glare points to be visible on the particles surface from an observation angle $theta$ (scattering angle). The glare point phenomenon can be easiest described, with the geometrical optics (GO) approximation, in which glare points can be represent the exiting points of the light rays on the surface of sphere. A light ray is either reflected at the surface of the sphere ($p=0$) or enters the sphere and exits it eventually after $p-1$ internal reflections (refracted ray $p\geq 1$), see Figure \ref{fig:IPI_GO_Sphere}. Due to different refraction at the particle surface (Snell's law), the light paths of refracted rays differ for droplets and bubbles (droplets having a larger refractive index $n_1$ than the surrounding medium $n_2$, $n_1>n_2$ applies, whereas for bubbles $n_1<n_2$ applies), see Figure \ref{fig:IPI_GO_Sphere}. The relation between reflective index of the sphere and its surrounding is described by the real part of the relative reflective index $m=n_1/n_2$. Note that the complex part of $m$, describing the dampening of the electromagnetic wave is not considered in the GO approximation. Using ray tracing, the relative glare point position $w$ on the particle can be calculated for a known scattering angle, see Figure \ref{fig:IPI_WorkingPrinciple}. Knowing the relative glare point position, the particle size can be calculated by determining the physical separation of two glare points $\Delta_{GP}$. When sizing (and tracking) a large number of particles in a large field of view (e.g in sprays), the individual particle image is to small to determine the glare point separation with sufficient accuracy. A solution to this problem, presents defocusing in IPI. Assuming under a certain scattering angle, only two glare points are visible and are equal in intensity, the two glare points act similar to a double slit in a Young's fringe experiment \cite{Semidetnov2003}, compare Figure \ref{fig:IPI_WorkingPrinciple}. This analogy to young's fringe experiment is essential for the working principle of the IPI method. Note that the equal intensity of two glare points is only provided at certain scattering angles, being the main limitation of scattering angles for IPI. Defocusing the particle image, the images of the two glare points overlap until eventually collapsing with sufficient distance from the focal plane. In this overlapping region an interference pattern is formed, see Figure \ref{fig:IPI_WorkingPrinciple}. Similar to Young's fringe experiment the glare point separation ('slit spacing') can then be derived from the fringe spacing of the interference pattern. This allows for high accuracy sizing of a large number of small particles with in an image.\\

In the forward scatter region, scattering angles of $\theta=$66° for water droplets ($m=1.333$) and $\theta=$45° for air bubbles in water ($m=1/1.333$) have been proven to work well for IPI. In this region the reflected $p=0$ and first order refracted $p=1$ glare points are equal in intensity. To calculate the glare point positions, first the incident angle of each ray relative to the sphere surface need to be calculated with respect to the scattering angle. Note that the complementary angle is used here for conformity with the commonly used convention for IPI, see Figure \ref{fig:IPI_WorkingPrinciple}. The complementary angle $\beta_r^{(0)}$ of the incident angle of the reflected ray ($p=0)$ is calculated straight forward from geometrical considerations:
\begin{equation}
    2\beta_r^{(0)}=\theta
\end{equation}
For the refracted ray ($p=1$) the angle of the incident ray $\beta_i^{(1)}$ is calculated from the total deflection $\theta'$ and the angle of the transmitted ray $\beta_t^{(1)}$. \cite{vandeHulst.1991}.
\begin{equation}
    \theta'=2(\beta_t^{(1)}-\beta_i^{(1)})=\theta
    \label{Equ:TotalDeflec_p1}
\end{equation}
The missing relation for $\beta_t^{(1)}$ is given by Snell's law:
\begin{equation}
    \cos{\beta_i^{(1)}}=m\,\cos{\beta_t^{(1)}}
    \label{Equ:Snell}
\end{equation}
The relative position $w$ of each glare point then calculates as a function of the particle radius $a$ and $q=\pm1$ depending on which side of the bubble the glare point is positioned, see Figure \ref{fig:IPI_WorkingPrinciple}. More detail on the determination of the sign of $q$ is presented in Section \ref{Sec:Visibility_Sub_GO} and Table \ref{tab: k_and_q_vaules}.
\begin{equation}
    w^*_0=a\,w_0=a\,q\cos{(\beta_r^{(0)})}
    \label{Equ:w_def}
\end{equation}
\begin{equation}
    w^*_1=a\,w_1=a\,q\cos{(\beta_i^{(1)})}
\end{equation}
These formulas provide the relation between the glare point separation and the particle size for the paring of the reflected and first order refracted glare point. Note that these formulas follow the convention of Van de Hulst \textit{et al.} \cite{vandeHulst.1991}, with $w$ being the glare point position relative to the particle radius $a$. Introducing the concept of Young's fringe experiment, the relation between number of stripes in the interference pattern and particle diameter can be drawn over the glare point separation. For droplets the relations is \cite{Roth1994,Rousselle1999}:
\begin{equation}
    2a=\frac{2\lambda N}{\alpha}(\cos{(\theta/2)}\,+\,\frac{m\sin{(\theta/2)}}{\sqrt{m^2-2m\cos{(\theta/2)}+1}})^{-1}
    \label{Equ:IPI_classic_drop}
\end{equation}
with $\lambda$ being the wave length of the illuminating light, $\alpha$ being the collecting angle of the imaging optics and $N$ being the number of fringes observable in the particle image. A similar relation results for bubbles\cite{Niwa.2000,Kawaguchi.2002}: 
\begin{equation}
    2a=\frac{2\lambda N}{\alpha}(\cos{(\theta/2)}\,-\,\frac{m\sin{(\theta/2)}}{\sqrt{m^2-2m\cos{(\theta/2)}+1}})^{-1}
    \label{Equ:IPI_classic_bubble}
\end{equation}
Equation \ref{Equ:IPI_classic_drop} and \ref{Equ:IPI_classic_bubble} link the angular fringe frequency to the glare point spacing. Since these formulas use the $p=0$ and $p=1$ glare point paring and the GO approximation, Equation \ref{Equ:IPI_classic_drop} for droplets is limited to scattering angles of $\theta=2\arccos{1/m}$ and Equation \ref{Equ:IPI_classic_bubble} for bubbles to $\theta=2\arccos{m}$ respectively. The limitation for droplets is a result of the incident angle of the $p=1$ ray being limited to entering the particle at an angle lower than 90° to the surface's orthogonal. This condition of the ray being able to enter the sphere results in the condition $\beta_i^{(1)}\geq0$. For the bubbles the limitation is the exiting condition of the $p=1$ ray, as the light ray needs to hit the surface at an exiting angle smaller than 90° from the surface orthogonal, to avoid total internal reflection. This results in the condition $\beta_t^{(1)}\geq0$. Note that at this scattering angle, the $p=0$ and $p=1$ glare point of the bubble collapse into a single point, providing no longer a second glare point for interference.
Consequently applying IPI with the $p=0$ and $p=1$ glare point paring is limited to the forward scatter region (at most $\theta=$90° due to the $\arccos$ term).\\

Recalling the analogy to Young's fringe experiment, the particle can be viewed as a black box with the glare point being point sources for the emission of light. The particle diameter can still be calculated as long as the glare point separation (slit spacing) is known. Therefore a more abstract formulation can be found. The fringe frequency $F$ of the interference pattern in some distance $S$ to the particle (the light traveling through a medium with reflective index $n_2$) is given by \cite{born_wolf_2019}:
\begin{equation}
    1/F=\frac{\lambda S/n_2}{\Delta^*_{\mathrm{GP}}}
    \label{Equ:YoungsFringe}
\end{equation}
with $\Delta*_{\mathrm{GP}}$ being the glare point separation of a chosen glare point paring. The term $\Delta^*_{GP}$ can be expressed as:
\begin{equation}
    \Delta^{*(i,j)}_{\mathrm{GP}\,(\theta)}=|w^*_{p_i}-w^*_{p_j}|=a\,|w_{p_i}-w_{p_j}|=a\,\Delta^{(i,j)}_{\mathrm{GP}\,(\theta)}
\end{equation}
for any paring of glare point $p=i$ and $p=j$, compare Equation \ref{Equ:w_def}. With Shen \textit{et al.} \cite{Shen.2012} providing a generalization to arbitrary optical systems along the optical axis, using the ray transfer matrix model, Equation \ref{Equ:YoungsFringe} can be generalized to:
\begin{equation}
    1/F=\left|\frac{\lambda B_{tot}}{\Delta^*_{\mathrm{GP}}}\right|
    \label{Equ:Sehn2012}
\end{equation}
The optical system along the optical axis is here described by the ray transfer matrix $M_{tot}$ with
\begin{equation}
    M_{tot}=\left[ \begin{array}{cc}
        A_{tot} & B_{tot} \\
        C_{tot} & D_{tot}
    \end{array}\right]
\end{equation}
Consequently the particle diameter for an arbitrary glare point paring and optical system is given by:
\begin{equation}
    2a=\left|\frac{\lambda B_{tot}}{\Delta^{(i,j)}_{\mathrm{GP}\,(\theta)}}\right|\cdot F
    \label{Equ:IPI_formula_general}
\end{equation}
Note that for a given wave length, scattering angle and optical system, the particle diameter is a linear function of the fringe frequency. In this formulation, the particle is considered as a black box with the glare points being point emitters of light. Equation \ref{Equ:IPI_formula_general} is valid in the front-, side- and back-scatter region (division by zero has to be avoided for certain glare point parings). \\
The glare point spacing $\Delta^{(i,j)}_{\mathrm{GP}\,(\theta)}$ is the only unknown for measurements in side- and back-scatter region. Therefore, in the first step applicable scattering angles $\theta$ need to be identified and the corresponding glare point paring $(i,j)$ has to be determined, which will be shown in Section \ref{Sec:IPIAngles}. Then the glare point separation can be calculated from geometrical relations.

\section*{Light Scattering at small Particles: Mie Theory and Debye Series Expansion}
\label{Sec:Mie}

For the identification of applicable scattering angles, a look inside the 'black box' particle must be taken. In general the scattering of light at spherical particles can be described by the Lorentz-Mie-theory \cite{vandeHulst1957}. The scattered light waves in Mie-theory can be described by the already discussed complex refractive index $m$ of the sphere (here the dampening of the light waves is considered in $\Im\{m\}$) and the size factor $x=2\pi a/(\lambda_0/n_2)$ of the sphere. With $\lambda_0$ being the wave length of the light in vacuum and $n_2$ the reflective index of the surrounding medium. Note that due to the change of the wave length of the light in the surrounding medium, an air bubble in water has a different size factor $x$ compared to a water droplet in air with the same diameter. The complex amplitude to the scattered light is given by \cite{vandeHulst1957}:
\begin{equation}
    S_{1\,(\theta)} = \sum_{n=1}^{\infty} \frac{2n+1}{n(n+1)}(a_n \Pi_{n\,(\theta)} + b_n\tau_{n\,(\theta)})
    \label{Equ:Mie_S1}
\end{equation}.
\begin{equation}
    S_{2\,(\theta)} = \sum_{n=1}^{\infty} \frac{2n+1}{n(n+1)}(b_n \Pi_{n\,(\theta)} + a_n\tau_{n\,(\theta)})
    \label{Equ:Mie_S2}
\end{equation}
from which the light intensity $I=\sqrt{S^2}$ and the phase of the wave can be directly calculated. In Equation \ref{Equ:Mie_S1} and \ref{Equ:Mie_S2}, $a_n$ and $b_n$ are the Mie coefficients corresponding to TM-waves (parallel polarization) and TE-waves (perpendicular polarization) respectively, which are functions of $m$ and $x$. The Mie angular functions $\Pi_{n\,(\theta)}$ and $\tau_{n\,(\theta)}$ are functions of the scattering angle and can be calculated from the Legendre functions of first kind ($k=1$) and $n^{\mathrm{th}}$ order (also called associated Legendre polynomials)\cite{vandeHulst1957}:
\begin{equation}
    \Pi_{n (\theta)}\,=\,\frac{P_{n\,(\cos(\theta))}^{(1)}}{\sin(\theta)}
    \label{Equ:Pie}
\end{equation}
\begin{equation}
    \tau_{n (\theta)}\,=\,\frac{\mathrm{d}}{\mathrm{d}\theta}P_{n\,(\cos(\theta))}^{(1)}
    \label{Equ:Tau}
\end{equation}
The Legendre functions can be calculated from the Legendre polynomials\cite{Courant1953}:
\begin{equation}
    P_{n\,(z)}^{(k)}\,=\,(-1)^k(1-z^2)^{k/2}\frac{\mathrm{d}^k}{\mathrm{d}z^k}P_{n(x)}
    \label{Equ:LegendreP}
\end{equation}
The Legendre polynomials of order $n$ can be expressed in a compact manner by Rodrigues formula \cite{Abramowitz1965}:
\begin{equation}
    P_{n(z)}\,=\,\frac{1}{2^nn!}\frac{d^n}{dz^n}(z^2-1)^n
    \label{Equ: Legendre Rodrigues}
\end{equation}
For large spheres $x\gg 1$, large partial waves $n\gg 1$ and $\theta\neq\ 0,180$, $\Pi_{n (\theta)}\ll \tau_{n (\theta)}$ applies and, therefore, $S_1$ becomes associated with the TE-mode and $S_2$ with the TM-mode \cite{vandeHulst1957}. For numerical calculation the using $n_{\mathrm{max}} \approx x + 4x^{1/3} + 2$ number of partial waves has found to be sufficient \cite{Bohren1998b}.\\ 

\begin{figure}
    \centering
    \subfigure[Defocusing glare points]{\includegraphics[width=0.4\textwidth]{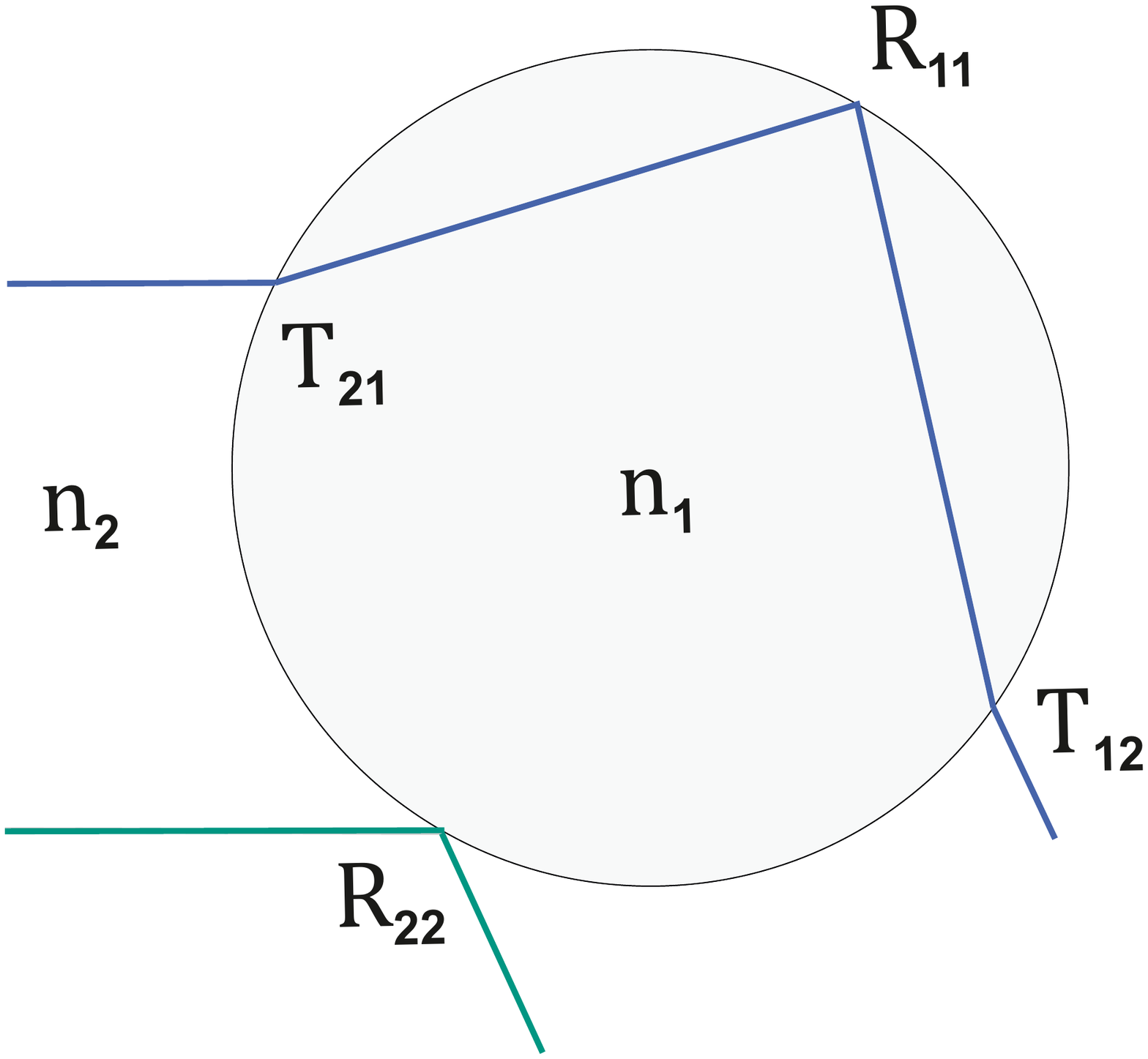}}
    \caption{Debye series expansion of the Mie coefficients visualized: The second term corresponding to the reflected wave $p=0$ ($\textcolor{mygreen}{\bullet}$) is described by the wave being reflected from medium 2 back to medium 2 (surrounding medium - $n_2$). The third term describes waves $p\geq1$ that are transmitted into the sphere (medium $2 \rightarrow 1$) and $p-1$ times internally reflected ($1 \rightarrow 1$) till finally transmitted out of the sphere again ($1 \rightarrow 2$), here shown on the example of an $p=2$ wave ($\textcolor{myblue}{\bullet}$) in the case of $m<1$.}
    \label{fig:Debye_explained}
\end{figure}

While Mie-theory lays the basics for the calculation of light scattering at small particles, it provides only integral information for the scattered light (no information on the scattering order $p$), which is however, needed for the present problem. This additional information can be obtained by describing the Mie coefficients with the Debye-series expansion of the Mie-theory \cite{Hovenac1992,Gouesbet2003}:
\begin{equation}
     \left[ \begin{array}{cc} a_n \\
     b_n\end{array}\right] =\frac{1}{2}\left(1\,-\,R_{n\,a,b}^{22}\,-\sum_{p=1}^{\infty}T_{n\,a,b}^{21}(R_{n\,a,b}^{11})^{p-1}T_{n\,a,b}^{12}\right)
    \label{Equ: Debye-MieCoeff}
\end{equation}
The Mie coefficients in Equation \ref{Equ: Debye-MieCoeff} are divided in three terms which can be interpreted as follows:
The first term $1$ describes the Fraunhofer diffraction around the sphere. The second term $R_{n\,a,b}^{22}$ describes waves reflected at the sphere surface ($p=0$). The third term $\sum_{p=1}^{\infty}T_{n\,a,b}^{21}(R_{n\,a,b}^{11})^{p-1}T_{n\,a,b}^{12}$ depicts the refracted waves ($p\geq 1$), which are first transmitted in the sphere ($T_{n\,a,b}^{21}$) then $p-1$ times internally reflected ($R_{n\,a,b}^{11})^{p-1}$) and finally transmitted out of the sphere ($T_{n\,a,b}^{12}$)\cite{Gouesbet2003}, see Figure \ref{fig:Debye_explained}. 
The reflection and transmission coefficients $R$ and $T$ respectively, can be calculated for incoming waves from the Riccati-Bessel functions $\xi_{n(x)}$ and $\Gamma_{n(x)}$ and their derivatives with respect to $x$ $\xi'_{n(x)}$ and $\Gamma'_{n(x)}$. \cite{Gouesbet2003}:
\begin{equation}
     R_{n\,a,b}^{22}\,=\,\frac{1}{D_{a,b}}[A\Gamma_{n(x)}'\Gamma_{n(y)}\,-\,B\Gamma_{n(x)}\Gamma_{n(y)}']
    \label{Equ: R22}
\end{equation}
\begin{equation}
     T_{n\,a,b}^{21}\,=\,2i\frac{B}{D_{a,b}}
    \label{Equ: T21}
\end{equation}
With $D_{a,b}$ being:
\begin{equation}
     D_{a,b}\,=\, -A\xi_{n(x)}'\Gamma_{n(y)}\,+\,B\xi_{n(x)}\Gamma_{n(y)}'
    \label{Equ: Debye D}
\end{equation}
With the following abbreviations being used for better readability: $y=x m$ and 
\begin{equation}
     \begin{split}A\,=\,m \,;\,B\,=\,1\,\, \mathrm{for\,\,TM-wave\,\,}a_n \\\
     A\,=\,1 \,;\,B\,=\,m\,\, \mathrm{for\,\,TE-wave\,\,}b_n \end{split}
    \label{Equ: Debye AB}
\end{equation}
For out-going waves the calculation follows accordingly \cite{Gouesbet2003}:
\begin{equation}
     R_{n\,a,b}^{11}\,=\,\frac{1}{D_{a,b}}[A\xi_{n(x)}'\xi_{n(y)}\,-\,B\xi_{n(x)}\xi_{n(y)}']
    \label{Equ: R11}
\end{equation}
\begin{equation}
     T_{n\,a,b}^{12}\,=\,2i\frac{A}{D_{a,b}}
    \label{Equ: T12}
\end{equation}
The Riccati-Bessel functions of third kind $\Gamma_{n(z)}$ and $\xi_{n(z)}$ are calculated from the Riccati-Bessel functions of first $\Psi_{n(z)}$ and second kind $\Omega_{n(z)}$, which can be further expressed as functions of the Bessel function of first $J_{n\,(z)}$ and second kind $Y_{n,(z)}$. For more detail on the Bessel function see e.g. \cite{Abramowitz1965}.  
\begin{equation}
     \left(
     \begin{array}{cc}
          \Gamma_{n(z)}  \\
          \xi_{n(z)}
     \end{array}\right) = \left(\begin{array}{cc}
          \Psi_{n(z)} + i\Omega_{n(z)}  \\
          \Psi_{n(z)} - i\Omega_{n(z)} 
     \end{array}\right) = \sqrt{\frac{\pi z}{2}} \left(\begin{array}{cc}
          J_{n+1/2\,(z)} + i Y_{n+1/2\,(z)}  \\
          J_{n+1/2\,(z)} - i Y_{n+1/2\,(z)} 
     \end{array}\right)
    \label{Equ: Riccati-Bessel}
\end{equation}
This lays the foundation for the numerical calculation of the scattering intensity and complex amplitude of each scattering order at a spherical particle for scattering angles from 0° to 180°.

\section*{Identification of applicable Scattering Angles}
\label{Sec:IPIAngles}

\subsection*{Derivation of the Visibility Formulation}
\label{Sec:Visibility}

With the calculation of the complex amplitude by means of the Debye series expansion, glare points on the particles surface can be investigated in greater detail to identify applicable scattering angles in the full range of $\theta=0..180$°. For the investigation, first boundary conditions for applicable angles need to be described. Recalling the analogy to Young's fringe experiment, the given IPI-framework, see Equation \ref{Equ:IPI_formula_general}, works as long as the analogy between the glare points on the particle and a double slit experiment is valid. In consequence, this analogy can be drawn if the glare points satisfy two conditions:\\

\begin{figure} [h]
\centering
    \subfigure{\includegraphics[width=0.4\textwidth]{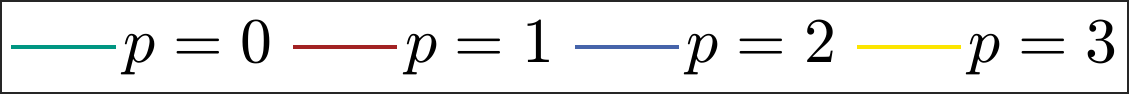}}\\
    \addtocounter{subfigure}{-1}
    \subfigure[Debye Plot - Droplet]{\includegraphics[width=0.49\textwidth]{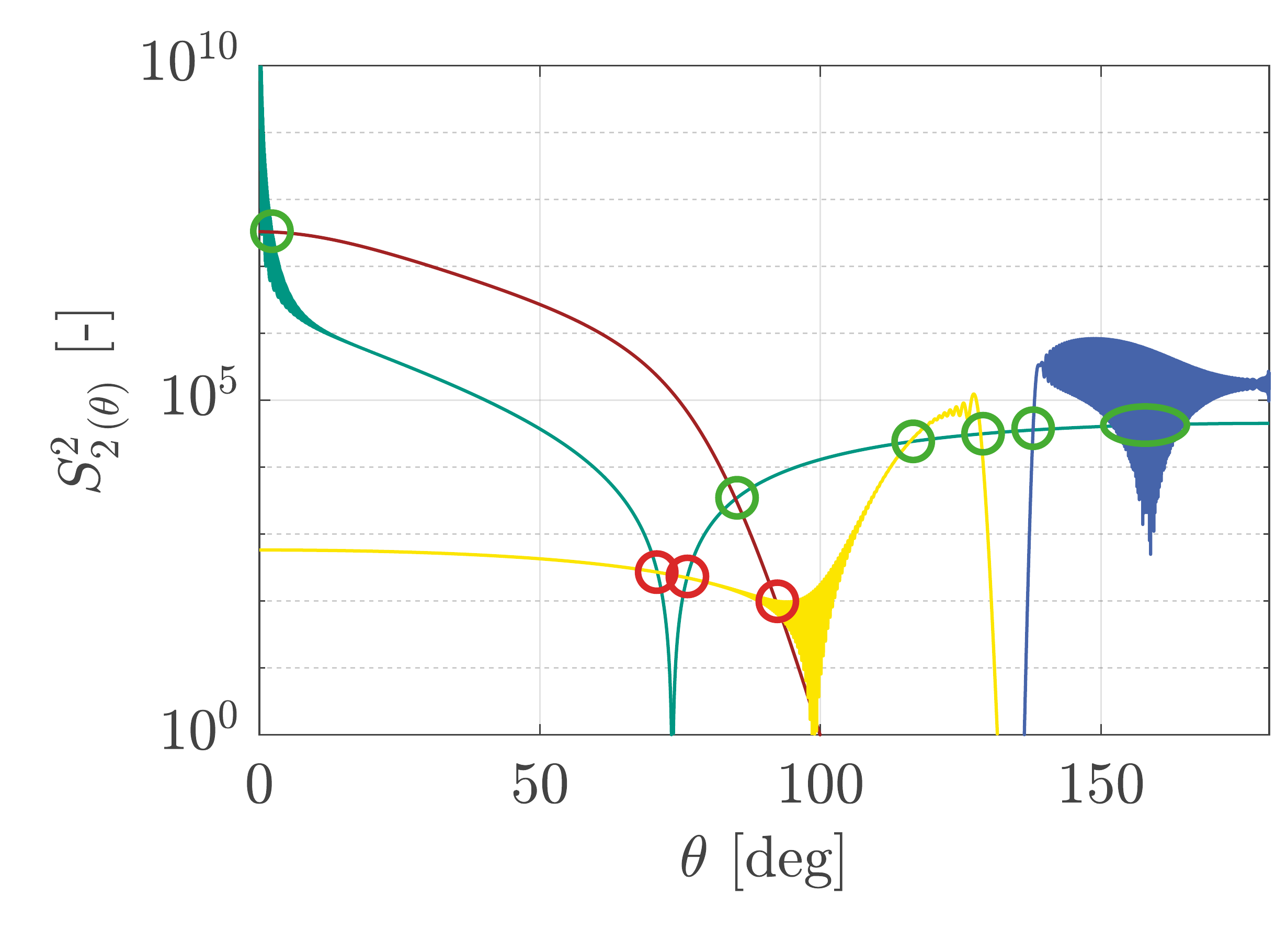}}
    \subfigure[Debye Plot - Bubble]{\hspace{1mm}\includegraphics[width=0.49\textwidth]{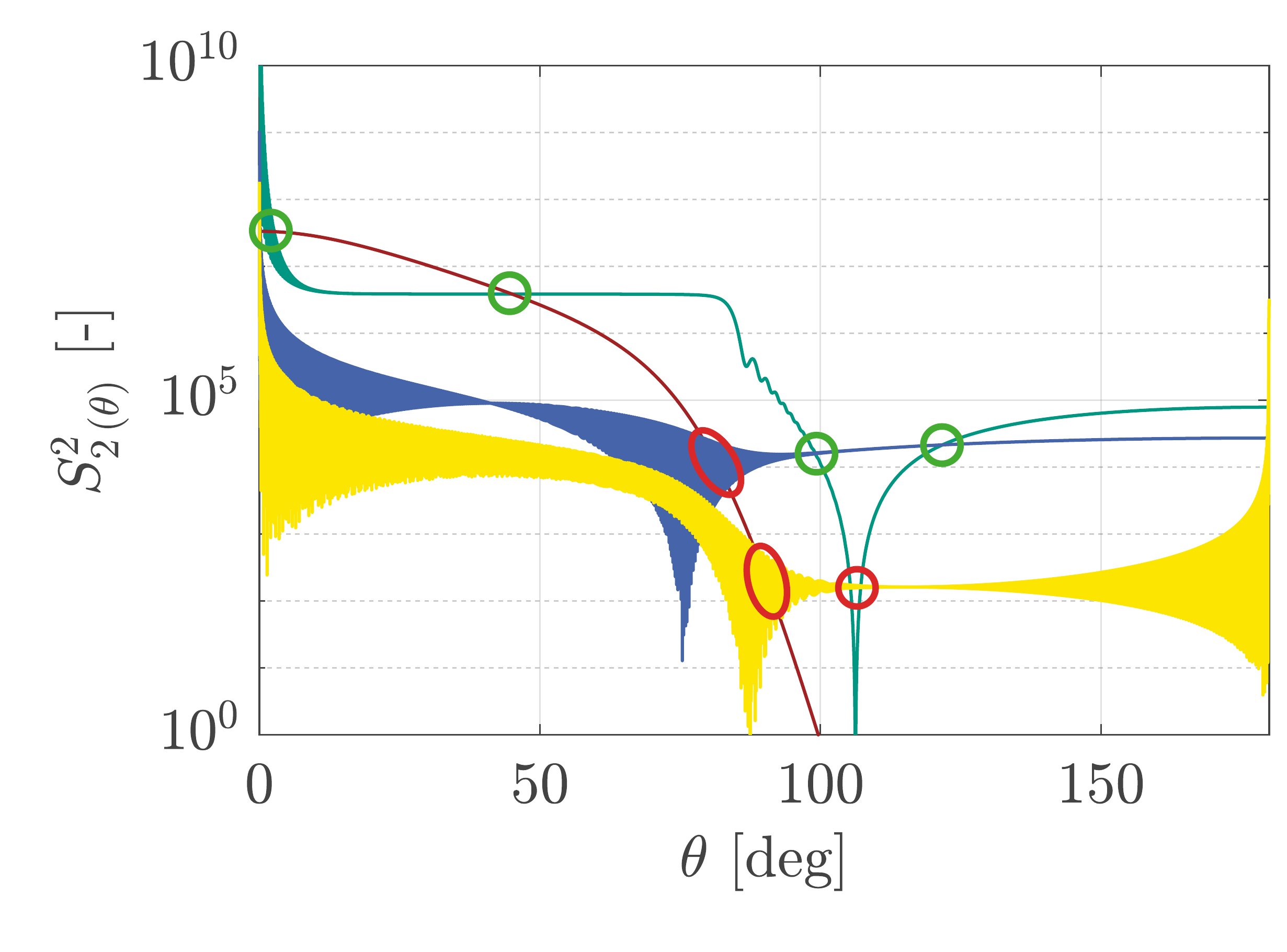}}
\caption{Scattering intensities of the TM-mode $S_2^2$ calculated from the Debye series expansion of Mie theory. Shown are Debye plots for a) water droplet in air $m=1.333+i1.82E-09$, $\lambda=532\mu m$, $a=250\mu m$ resulting in $x=2952.6$ and b) air bubble in water $m=1/1.333+i0$, $\lambda=532\mu m$, $a=250\mu m$ resulting in $x=3935.8$. The visibility criteria are visualized: Scattering angles satisfying both criteria are marked in green ($\textcolor{mygreen2}{\bullet}$), whereas angles only matching the first condition but not the second one are marked in red ($\textcolor{myred2}{\bullet}$)}
\label{fig:Debye_Conditions}
\end{figure}

\begin{figure} [h]
\centering\subfigure{\includegraphics[width=0.4\textwidth]{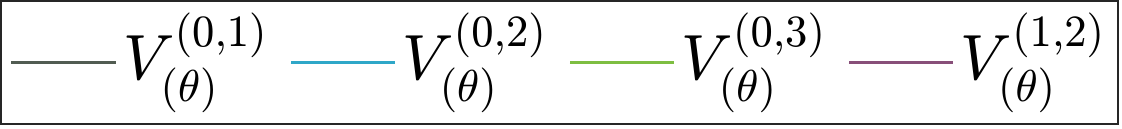}}\\
\addtocounter{subfigure}{-1}
    \subfigure[Debye Plot - Droplet - TE]{\includegraphics[trim={0 0 0 1.5cm},clip,width=0.49\textwidth]{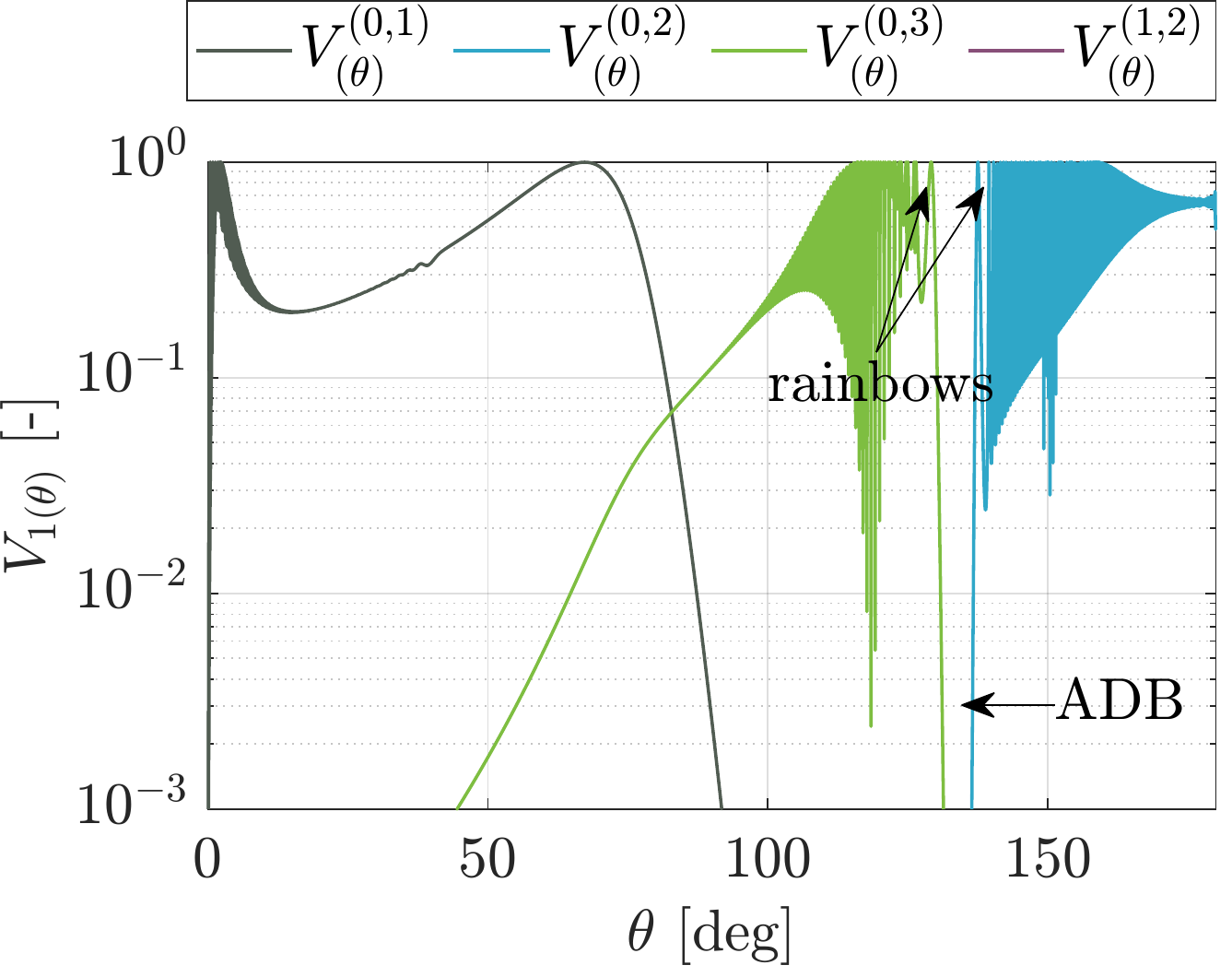}}
    \subfigure[Debye Plot - Droplet - TM]{\hspace{1mm}\includegraphics[trim={0 0 0 1.5cm},clip,width=0.49\textwidth]{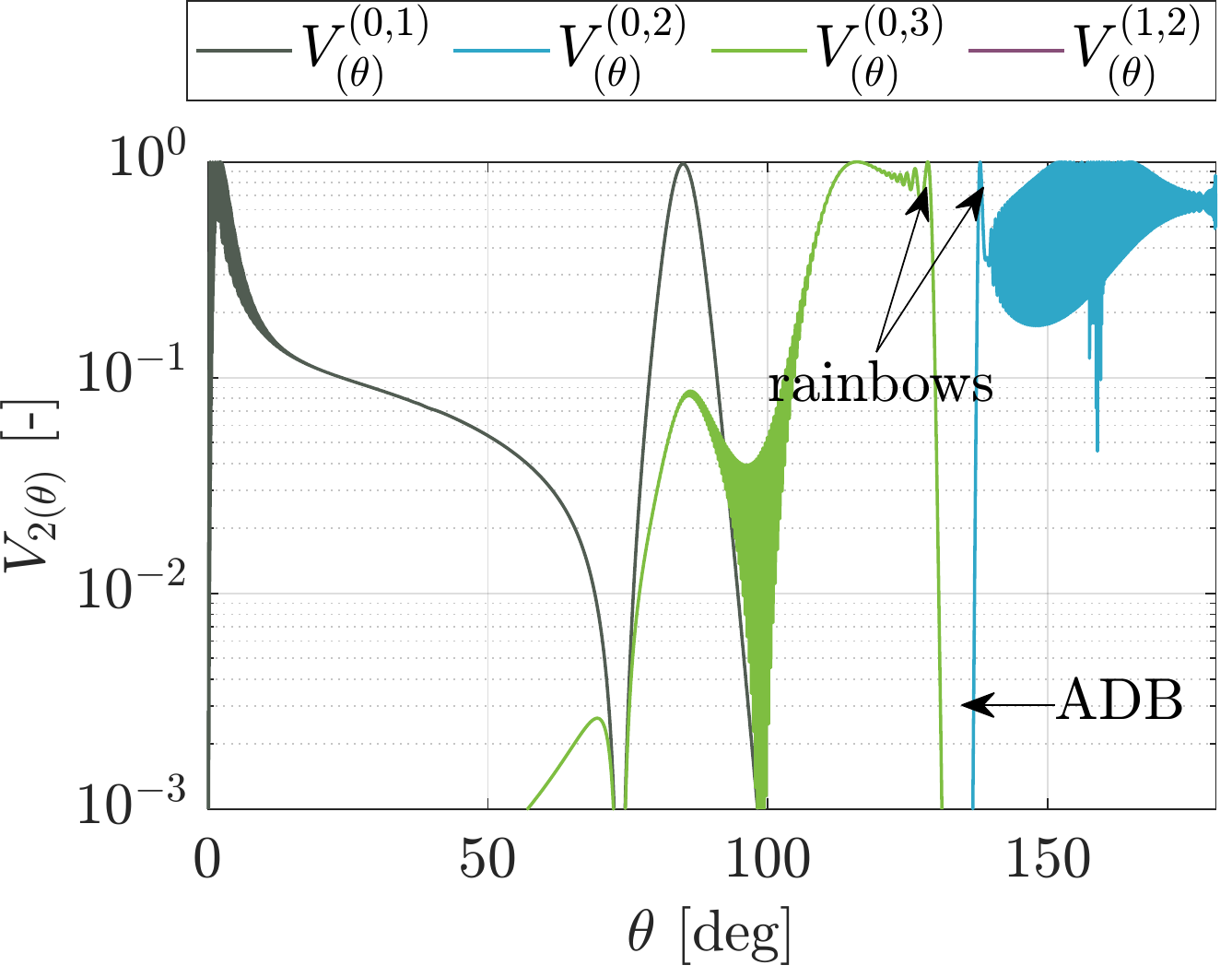}}
    \subfigure[Debye Plot - Bubble - TE]{\hspace{1mm}\includegraphics[trim={0 0 0 1.5cm},clip,width=0.49\textwidth]{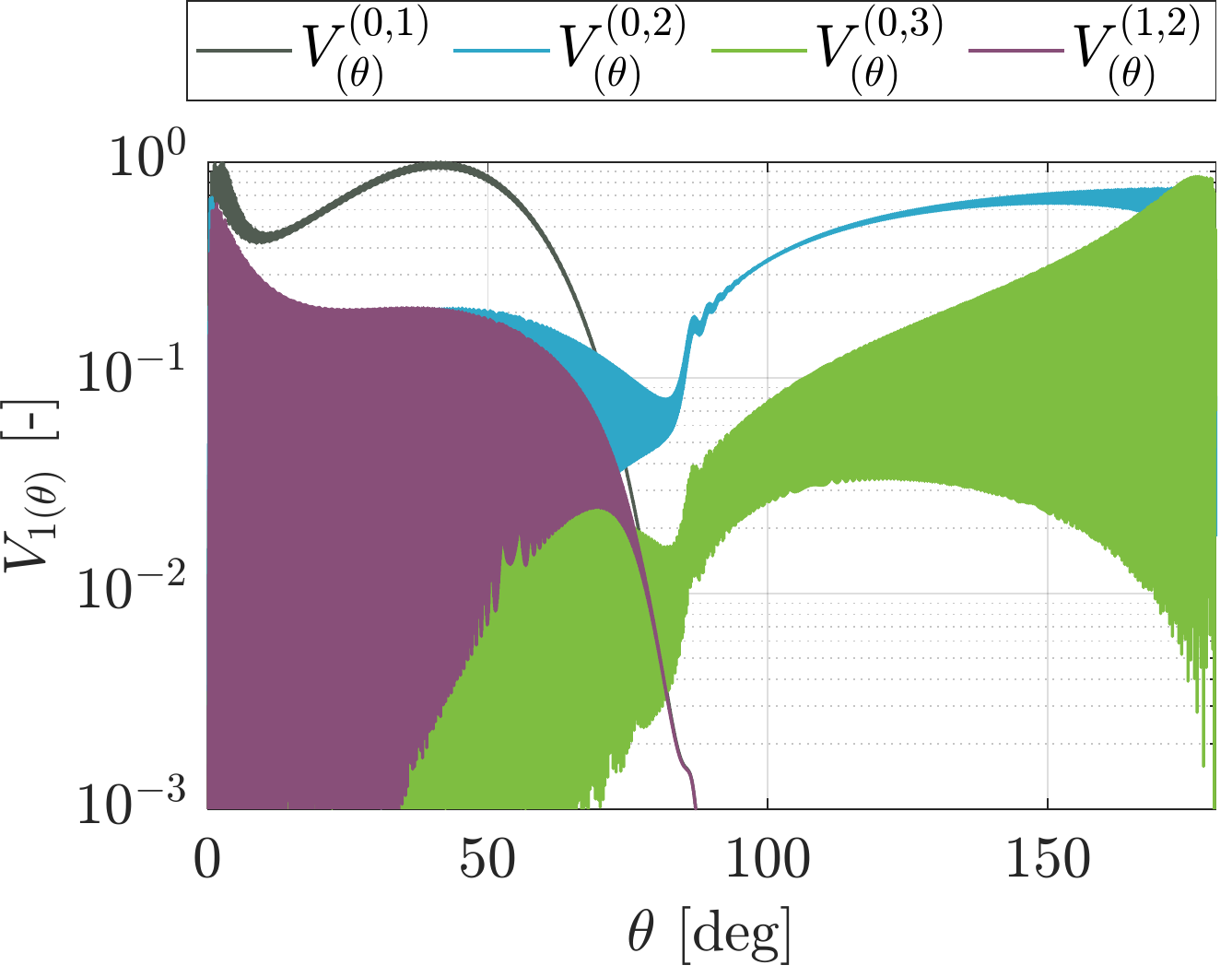}}
    \subfigure[Debye Plot - Bubble - TM]{\hspace{1mm}\includegraphics[trim={0 0 0 1.5cm},clip,width=0.49\textwidth]{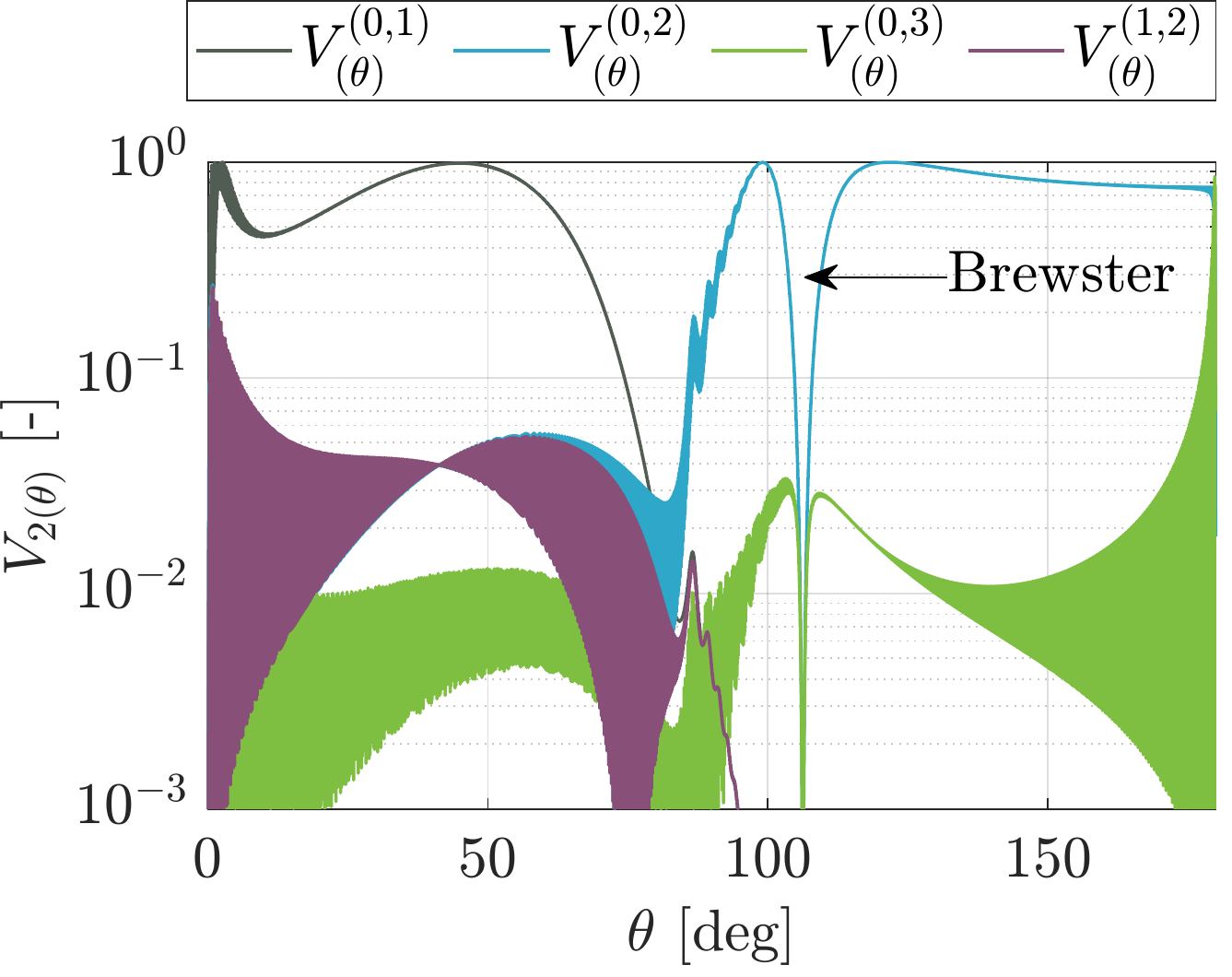}}
\caption{Visibility plot over the scattering angle for a water droplet in air $m=1.333+i1.82\cdot10^{-9}$, $\lambda=532\mu m$, $a=250\mu m$ resulting in $x=2952.6$ in a) TE and b) TM-mode and an air bubble in water $m=1/1.333+i0$, $\lambda=532\mu m$, $a=250\mu m$ resulting in $x=3935.8$ in c) TE and d) TM-mode. A sufficient scattering angle for IPI is given whenever the visibility plot has a value close to one. Shown are the glare point parings (0,1), (0,2), (0,3) and (1,2). Additionally the rainbow angles of the $p=2$ and $p=3$ scattering orders and Alexanders dark band (ADB) are marked in a) and b). The Brewster angle is marked in d)}
\label{fig:Visibility_plots}
\end{figure}

First considering only two glare points on a particle with intensities $I_{p_i}$ and $I_{p_j}$, to form a visible interference pattern, both glare points visible $p=i$ and $p=j$ must have approximately the same intensity (at least within the same order of magnitude). Otherwise no clearly visible intensity maxima and minima will be observable. This can be expressed in the form of of:
\begin{equation}
    \frac{I^{(p_i)}_{(\theta)}}{I^{(p_j)}_{(\theta)}} \approx 1
    \label{Equ: Visibility_first_Condition}
\end{equation}
with Equation \ref{Equ: Visibility_first_Condition} approaching 1 for a good visibility. Accounting for a phase difference of the interfering waves (due to the different path length traveled trough the particle, see Figure \ref{fig:IPI_GO_Sphere}) the visibility can be expressed as \cite{Dehaeck2008,Zhang.2018}:
\begin{equation}
    V_{\mathrm{nec}(\theta)}^{(p_i,p_j)} = \frac{2\sqrt{I_{(\theta)}^{p_i} I_{(\theta)}^{p_j}}}{I_{(\theta)}^{p_i} + I_{(\theta)}^{p_j}} \approx 1
    \label{Equ: Visibility_traditional}
\end{equation}
This is the necessary condition to form an interference pattern in the first instance. However, this condition ignores the presence of further glare points, which is sufficient for the front scatter regime as the $p=0$ and $p=1$ scattering orders dominate this region. This howbeit, not necessarily the case for other scattering angles. The intensity of the two relevant glare points (those which are equal in intensity - satisfying the first condition) must be the dominating glare points  for the given scattering angle to form a visible interference pattern. Therefore, a second sufficient condition is needed, ensuring that the chosen glare point pair ($p=i,j) $is not eclipsed by other brighter glare points $p\neq i,j$. This can be formulated as follows:
\begin{equation}
    I^{(p_i)}_{(\theta)} \&\, I^{(p_j)}_{(\theta)} \gg I^{p_{(k\neq i,j})}_{(\theta)}
    \label{Equ: Visibility_second_Condition}
\end{equation}
Reformulation Equation \ref{Equ: Visibility_second_Condition} in a similar manner to Equation \ref{Equ: Visibility_traditional}, the following expression is derived for the sufficient visibility criterion:
\begin{equation}
    V_{\mathrm{suf}(\theta)}^{(p_i,p_j)} = \frac{2\sqrt{I_{(\theta)}^{(p_i)} I_{(\theta)}^{(p_j)}}}{\sum_{k=0}^{\infty}I_{(\theta)}^{(p_k)}} \approx 1
    \label{Equ: Visibility_intensityratio}
\end{equation}
Equation \ref{Equ: Visibility_intensityratio} approaches one if the considered glare point pair comprise the majority of light intensity emitted from the particle. Note that the denominator is the light intensity according to the Mie calculation with out Debye series expansion (calculating the Mie-Intensity is less computationally expensive compared to a significant amount of individual scattering orders by means of Debye series expansion for the $\sum_{k=0}^{\infty}I_{(\theta)}^{(p_k)}$ term).
Both conditions are visualized in Figure \ref{fig:Debye_Conditions}.
Since both Equations \ref{Equ: Visibility_traditional} and \ref{Equ: Visibility_intensityratio} approach one when the underlying condition is satisfied, they can be combined by means of multiplication to form a generalized visibility formulation:
\begin{equation}
    V_{(\theta)}^{(p_i,p_j)} = \frac{4 I_{(\theta)}^{(p_i)} I_{(\theta)}^{(p_j)}}{(I_{(\theta)}^{(p_i)}+I_{(\theta)}^{(p_j)})\sum_{k=0}^{\infty}I_{(\theta)}^{(p_k)}}
    \label{Equ: Visibility general}
\end{equation}
This new formulation for the visibility allows for the identification of applicable scattering angles and the according glare point paring for IPI, that satisfy both visibility criteria. Following this formulation, two visibility plots for a water droplet in air and an air bubble in water are shown in Figure \ref{fig:Visibility_plots}. It shows that both for water droplet and air bubbles in water, there are applicable scattering angles for IPI not only in the front- but also in the side- and back-scatter regime. Furthermore, the visibility changes depending on the polarization (parallel/perpendicular), resulting in certain scattering angles only being sufficient for a certain polarization. 

\FloatBarrier

\subsection*{Investigation of Glare Points of Water Droplets}
\label{Sec:Visibility_Sub:GP_Drop}

The visibility plots provide information on which angle a scattering order is applicable for IPI measurements. In this section a closer look at six scattering angles for droplets will be taken. As it is probably the most common example, water droplets ($m=1.333+i1.82\cdot10^{-9}$) were chosen, but the method is also applicable for other fluids like oil droplets. For other reflective indices ($m>1$) the angles in which a visibility of one is approached will shift slightly, apart from that will the overall structure of the visibility plots remain similar. \\
Considering Figure \ref{fig:Visibility_plots} for water droplets scattering angles of $\theta=$66°, 84.5°, 116°, 129°, 137,9° and around 160° $\pm$ 5° achieve visibility of 100\%. Nonetheless, is the process not finished at this point, since a single scattering order can result in more than one glare point along the particle surface. This can be thought of multiple possible ray paths through the particle with the same number of internal reflection (GO approximation). As the visibility plot shows the integral intensity of each glare point order at a certain scattering angle, this up-splitting of glare points is not considered. Consequently, a more detailed investigation is necessary.

To examine the glare points at a certain angle in more detail, the scattered light intensity can be plotted over the particle surface ($w-coordinate$). These glare point plots are calculated in a similar fashion to the Mie calculations of Van de Hulst \textit{et al.} \cite{vandeHulst.1991}, however, instead of using the Mie intensity for the integral intensity, the complex amplitude from the Debye series expansion $S^{(p_i)}_{(\theta)}$is used. This way the scattering intensity of each individual order $p$ can be plotted over the particle surface. The particle surface is parameterized by the $w$ coordinate from $w=-1...1$ over the diameter of the projected particle area \cite{vandeHulst.1991}, see Figure \ref{fig:IPI_GO_Sphere}. \\
Using Fourier optics, the scattering intensity $A_{(\theta)}$ along $w$ can be calculated from the complex amplitude $S_{(\theta)}$ for a certain scattering angle range $\theta\pm\Delta\theta$ by Equation \ref{Equ: A_w}\cite{vandeHulst.1991}:
\begin{equation}
    A_{(\theta)} = \int^{\theta_0+\Delta\theta}_{\theta_0-\Delta\theta} S_{(\theta)}e^{-ixw\cdot(\theta_0-\theta)}d\theta
    \label{Equ: A_w}
\end{equation}
with $S_{1\,(\theta)}$ for $A_{1\,(\theta)}$ (TE-mode) and $S_{2\,(\theta)}$ for $A_{2\,(\theta)}$ (TM-mode). The considered $\theta$-range is given by the observation angle $\alpha$ of the given setup with $\Delta\theta=\mathrm{tan}(\alpha)$. Equation \ref{Equ: A_w} takes the form of a (short time) Fourier transform, and consequently, $A_{(\theta)}$ along $w$ is the Fourier transformed of $S_{(\theta)}$. Note the analogy to the frequency and time/space domain of the standard Fourier transform, with $A_{\,(\theta)}$ resembling the frequency, $\theta$ the time/space domain and $w$ the frequency domain. Similar to the uncertainty relation for the Fourier transform over limited spacial domain (Fourier trade-off), an uncertainty of $A_{\,(\theta)}$ (glare point position) and scattering angle exists, which can be quantified by \cite{vandeHulst.1991}::
\begin{equation}
    \Delta\theta\cdot \Delta w = 1/x
    \label{Equ: uncertainty}
\end{equation}
Consequently both the glare point position and scattering angle can not be known with infinite accuracy at the same time. Mathematically this can be understood as a broader integration interval leading to a better defined $A_{\,(\theta)}$ but the range of considered scattering angles is larger, hence more $\theta$ is more uncertain. According to Equation \ref{Equ: uncertainty} this uncertainty becomes larger for smaller particles, physically limiting IPI towards small particles. To provide a physical interpretation of the uncertainty, the glare points can be thought of turning from distinct points into larger less distinct patches, with their separation becoming blurry. Note that in theory there is no upper limit towards the maximum measurable particle size, since the Mie-theory has no upper limit for particle sizes. However, an upper limit is presented by the Nyquist frequency (closest distinguishable fringe distance) set by the image resolution (camera chip size and imaging lens), compare Equation \ref{Equ:IPI_formula_general}. 

\begin{figure} [h]
\centering

    \centering\subfigure{\includegraphics[width=0.4\textwidth]{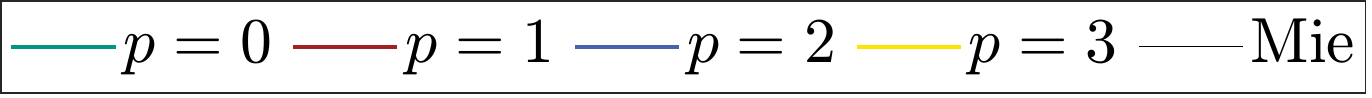}}\\
\addtocounter{subfigure}{-1}
    \subfigure[perpendicular - TE]{\includegraphics[width=0.49\textwidth]{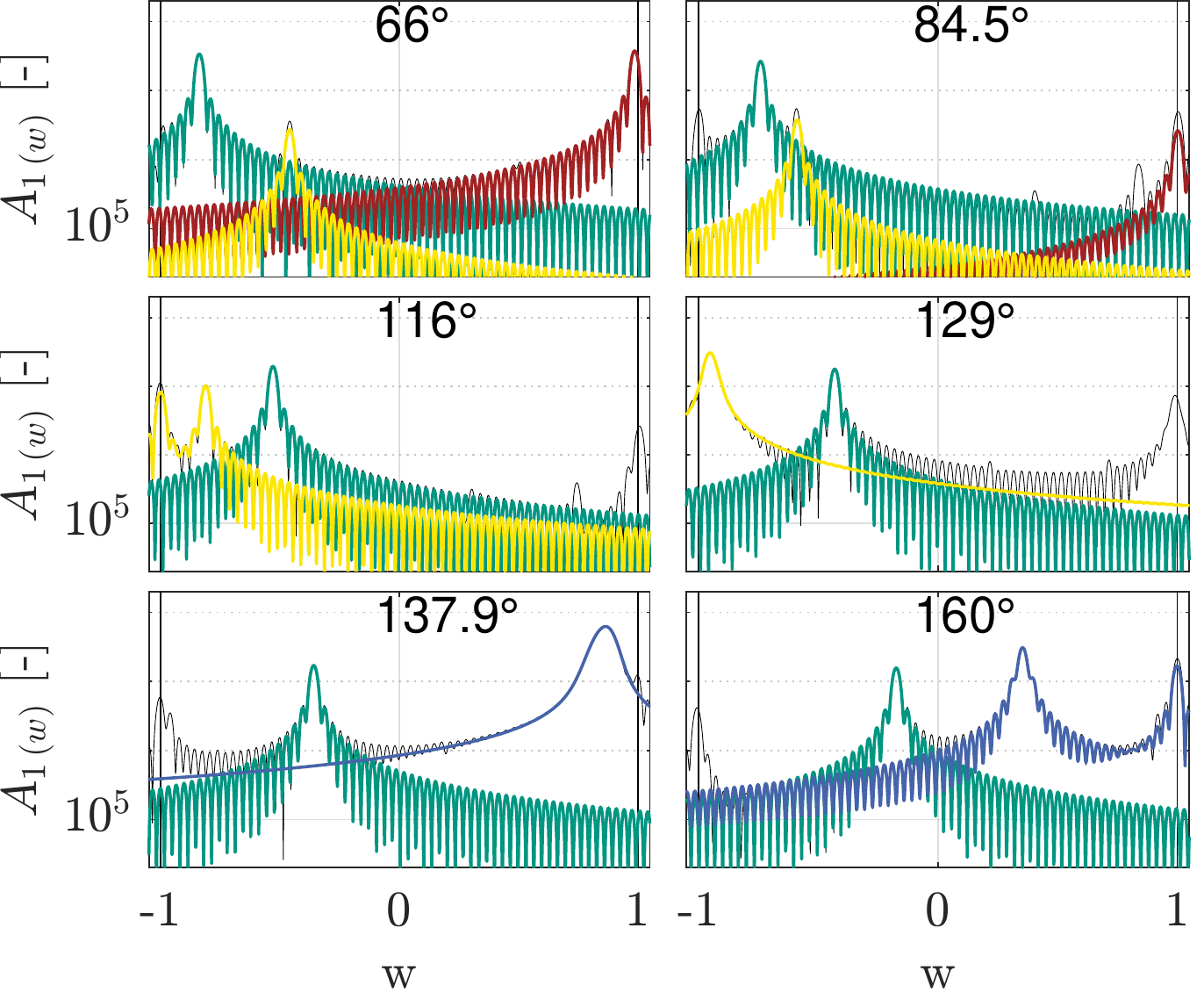}}
    \subfigure[parallel - TM]{\includegraphics[width=0.49\textwidth]{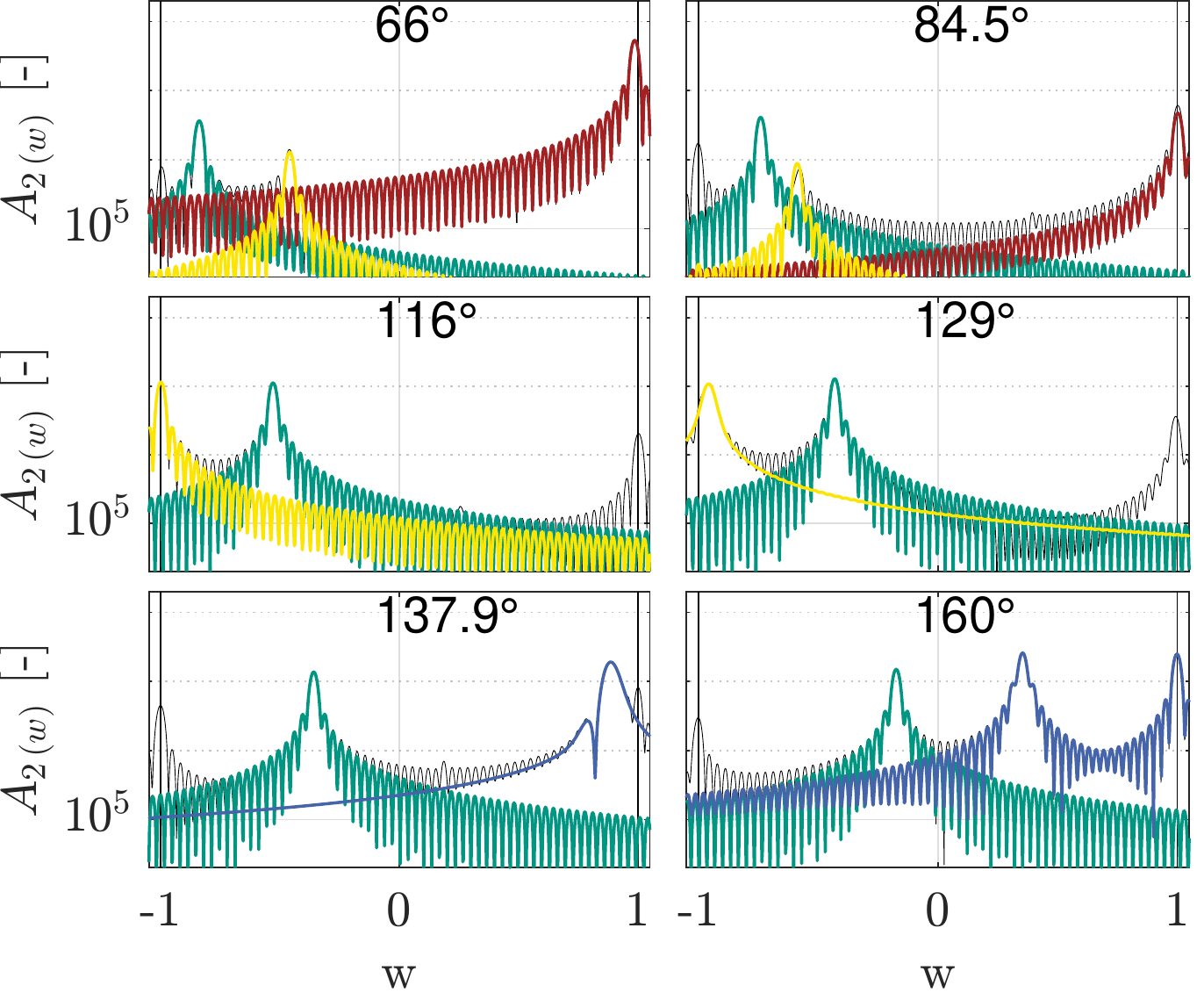}}
\caption{Glare Points by Debye series calculation visualized: The scattering intensity $A_{(w)}$ is plotted over the particle surface $w$ for a particle with $2a=250\,\mu m$, $\lambda=532\,nm$ (resulting in $x=2952.6$) and $m=1.333+i1.82\cdot10^{-9}$. The scattering uncertainty $\Delta \theta=3.7$ was chosen based on the optical system used for the validation experiment. The scattering angle $\theta_0$ around which $A_{(w)}$ was calculated is shown at the top of each graph. Depicted are the intensities of the scattering orders $p=0$ ($\textcolor{mygreen}{\bullet}$), $p=1$ ($\textcolor{myred}{\bullet}$), $p=2$ ($\textcolor{myblue}{\bullet}$) and $p=3$ ($\textcolor{myyellow}{\bullet}$), as well as the integral intensity of all scattering orders (Mie-intensity) ($\textcolor{myblack}{\bullet}$).}
\label{fig:GP_Droplet}
\end{figure}

Analyzing the glare points for a water droplet in Figure \ref{fig:GP_Droplet} at $\theta=66$° it can be seen that both for $A_{1\,(\theta)}$ (TE) and $A_{2\,(\theta)}$ (TM) the $p==$ and$p=1$ orders result in the highest intensity. However, for the TM-mode the glare point do not have equal intensity (see lower visibility, Figure \ref{fig:Visibility_plots}). Consequently for $\theta=66$° the TE-mode is better suited for IPI. For scattering angles of 84.5° the case is the opposite, in which the TM-mode achieves equal glare point intensity for the paring $p=(0,1)$. Moving to larger scattering angles, beyond the usually applied front scatter regime, the next applicable angle is $\theta=116$°. At this angle, the visibility plot suggests good visibility for the $p=(0,3)$ paring both for TE and TM-mode. It can be seen however, that for the TE-mode the $p=3$ scattering order splits up into two glare points. The integral intensity of the two $p=3$ glare points is equal to the $p=0$ glare point hence the high visibility. However, this three glare point configuration is not easily interpreted with Equation \ref{Equ:IPI_formula_general}, and hence not suitable. This demonstrates why the visibility plot should only be used a as precursor to find $\theta_0$. 
A closer investigation of $A_{(\theta)}$ is still needed. It should be noted that a third glare point can be utilized for redundancy \cite{Dehaeck2008}, but is beyond the scope of the present work. \\
The problem of the second $p=3$ glare point in the TE-mode, can be circumcised by choosing the rainbow angle of $p=3$. At the rainbow angle the two glare points of the same order collapse into a single glare point, see Figure \ref{fig:GP_Maps}. This is the case for $\theta=$129° and $p=(0,3)$. While a scattering angle of 129° is applicable both for TE and TM-mode, the TM-mode achieves a more similar intensity and the parasite glare points of other higher scattering orders also have lower intensity in this polarization. The same case can be observed for the $p=(0,2)$ paring which according to the visibility should result in a wide range of applicable scattering angles around 150°. However in this case also a second $p=2$ glare point exists. Choosing the rainbow angle for $p=2$ at $\theta=$137.9° the two $p=2$ glare points collapse into a single glare point. Consequently in back scatter a scattering angle of 137.4° is well suited for water droplets.  

While the rainbow angles of $p=2$ and $p=3$ do provide excellent visibility for interference patterns for IPI in the back-scatter regime, the angle range in between these two rainbow angles is not suitable for IPI, see Figure \ref{fig:Visibility_plots}. In the range of $\theta$=131°..136,6° (between the two rainbow angles) the visibility of the $p=(0,2)$ and $p=(0,3)$ paring drops close to zero. This is the angle range is called Alexanders dark band (ADB), in which the light intensity of the $p=2$ and $p=3$ drop close to zero, see Figure \ref{fig:Debye_Conditions} and Figure \ref{fig:GP_Maps}, hence the drop in visibility. Since the other higher scattering orders are also small in intensity, consequently, no sufficient interference patterns for IPI can be obtained in this range.

\subsection*{Investigation of Glare Points of Air Bubbles in Water}
\label{Sec:Visibility_Sub:GP_Bubble}

\begin{figure} [h]
\centering

    \centering\subfigure{\includegraphics[width=0.4\textwidth]{Figures/WaveOptics/GP_Bubbles/Legend.pdf}}\\
\addtocounter{subfigure}{-1}
    \subfigure[perpendicular - TE]{\includegraphics[width=0.49\textwidth]{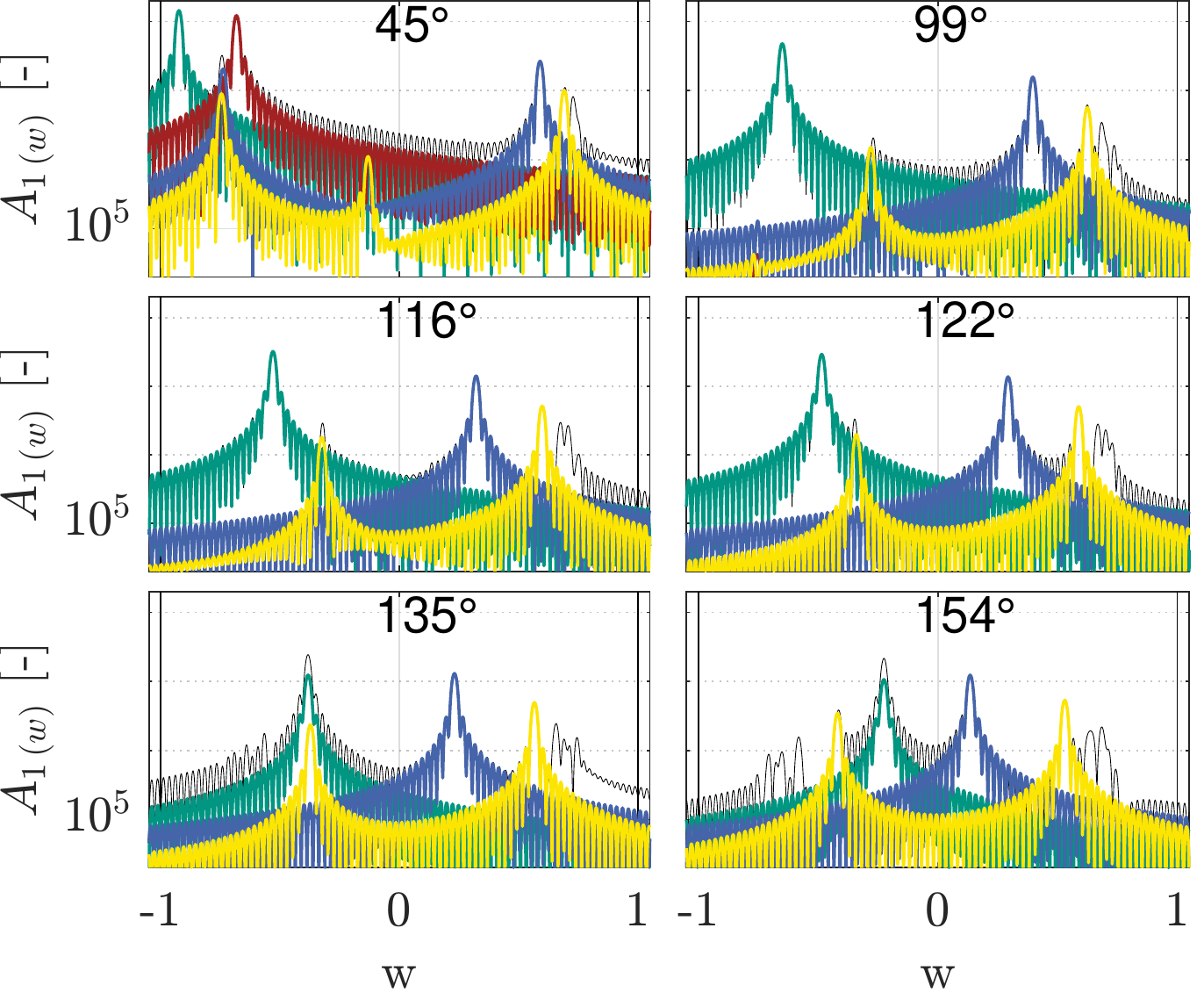}}
    \subfigure[parallel - TM]{\includegraphics[width=0.49\textwidth]{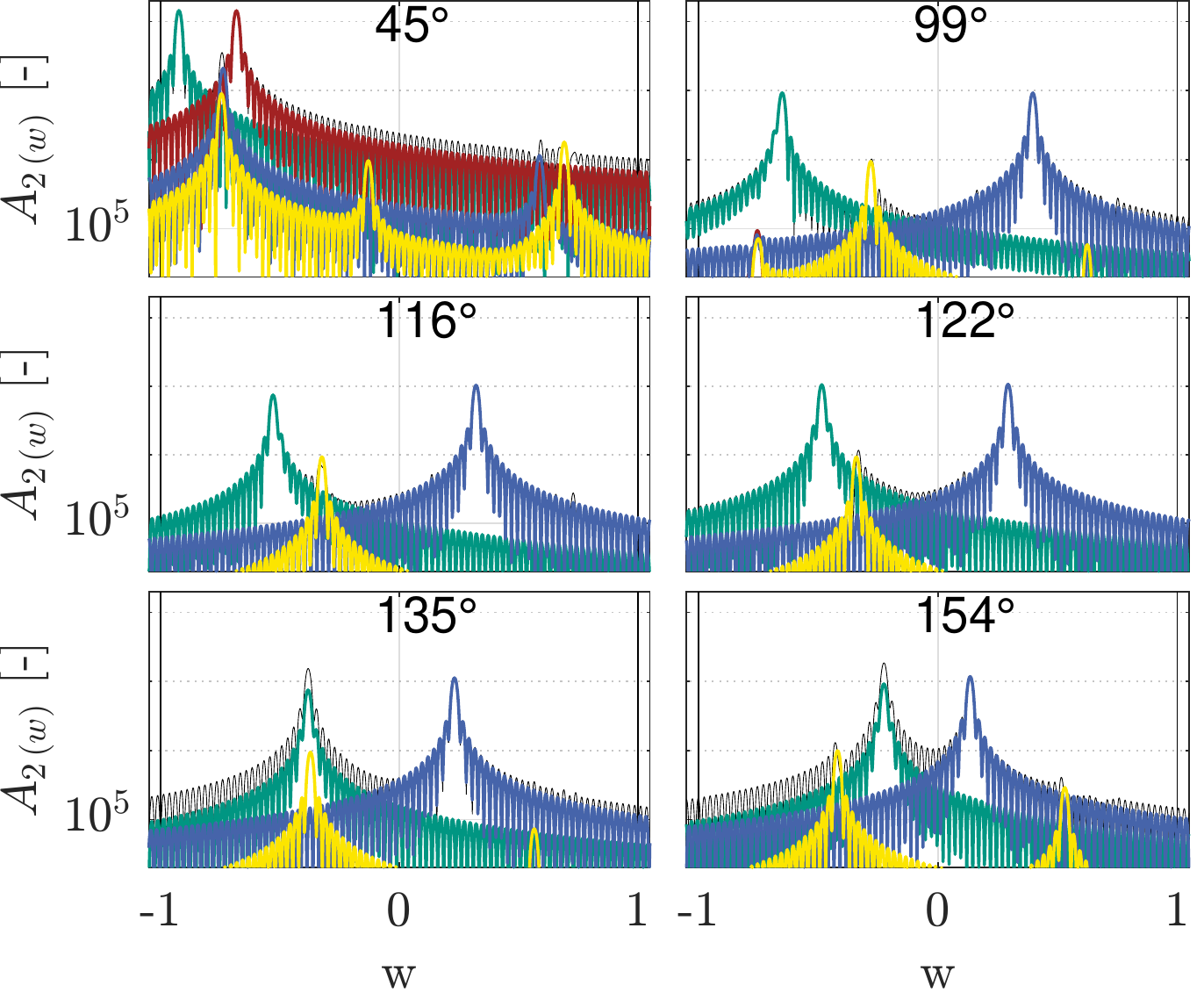}}
\caption{Glare Points by Debye series calculation visualized: The scattering intensity $A_{(w)}$ is plotted over the particle surface $w$ for a particle with $2a=250\,\mu m$, $\lambda=532\,nm$ (resulting in $x=3935.8$) and $m=1.333+i1.82\cdot10^{-9}$. The scattering uncertainty $\Delta \theta=3.7$ was chosen based on the optical system used for the validation experiment. The scattering angle $\theta_0$ around which $A_{(w)}$ was calculated is shown at the top of each graph. Depicted are the intensities of the scattering orders $p=0$ ($\textcolor{mygreen}{\bullet}$), $p=1$ ($\textcolor{myred}{\bullet}$), $p=2$ ($\textcolor{myblue}{\bullet}$) and $p=3$ ($\textcolor{myyellow}{\bullet}$), as well as the integral intensity of all scattering orders (Mie-intensity) ($\textcolor{myblack}{\bullet}$).}
\label{fig:GP_Bubble}
\end{figure}

Akin to the analysis for water droplets, in this section, air bubbles in water ($m=1/1.333$) will be analyzed, as they represent the most common case for applications with $m<1$ in IPI. Similar to the droplets are the exact angles dependent on the relative reflective index, but the basic characteristics distinguishing the bubble from the droplet remain the same for $m<1$. The most commonly used scattering angle for water bubbles is $\theta=$45° with the $p=(0,1)$ paring, which proves to be applicable both for TE and TM-mode. Note however that for the TE-mode the parasite glare points of $p=2$ and $p=3$ diminish in the background intensity and, therefore, the TE-mode is better suited, see Figure \ref{fig:GP_Bubble}. Moving to higher scattering angles in the side- and back-scatter region, the dominant glare point paring is $p=(0,2)$. As can be see in Figure \ref{fig:Visibility_plots}, the visibility of the $p=(0,2)$ paring in TE-mode is sufficient for almost all scattering angles in the back scatter regime, with the visibility increasing towards higher scattering angles. This is a desirable characteristic as scattering angles closer to 180° allow for single optical access applications. The TM-mode achieves even higher visibility in the back-scatter regime, but has a steep drop in visibility between $\theta=$99° and 116°, which should be avoided for sufficient IPI measurements. While the $p=3$ order is split up in multiple glare points, the dominant orders $p=(0,2)$ do not split up, so rainbow angles do not need to be considered for this case. For the visualization of the glare points different scattering angles 116° ($V^{(0,2)}=90\%$), 122° ($V^{(0,2)}=100\%$) 135° ($V^{(0,2)}=90\%$) and 154° ($V^{(0,2)}=80\%$) in TM mode are depicted to test their applicability. It shows that for all these angles the TM-mode is better suited compared to the TE-mode since the $p=(0,2)$ glare points are more similar in intensity and other higher order parasite glare glare points are less distinct. Consequently, a wide variety of scattering angles with the $p=(0,2)$ paring is applicable in the back-scatter regime for bubbles, preferably choosing the TM-polarization. 

While for the TM-mode, the scattering angles of 99° and 116° produce interference patterns close to 100\% visibility, there can be no sufficient interference patterns obtained in between these angles for scattering angles close to $\theta=$106°. In Figure \ref{fig:Visibility_plots} a steep drop of the visibility of the $p=(0,2)$ and $p=(0,2)$ parings close to zero can be noticed. Using the Fresnel equations \cite{Nolting2013}, it can be seen that around this scattering angle, the reflection coefficient becomes close to zero for the TM-mode (Brewster angle and its proximity). Consequently there is no reflected ($p=0$) light visible from this angle since all light is transmitted into the bubble, note the steep decrease of the $p=0$ intensity in Figure \ref{fig:Debye_Conditions}. Hence, no glare point paring ($p=(0,j)$) with the $p=0$ order glare point can produce a visible interference pattern at this angle. Since the other higher scattering orders have only little intensity in this range, in general no sufficient interference patterns can be obtained for bubbles at the Brewster angle and its proximity ($\theta=$106°) in the TM-mode. Note that the reflection coefficient is not zero for the TE-mode, for which applicable interference patterns can be obtained in this region. 

Furthermore for bubbles in the back-scatter regime, the glare point paring $p=(0,2)$ dominates for all scattering angles in the range $116°<\theta<180°$ (close to 180° the $p=3$ order also becomes relevant making this case more complex). As can be seen in Figure \ref{fig:GP_Bubble} and \ref{fig:GP_Maps} the $p=0$ and $p=2$ glare points move closer together with increasing scattering angles. Since the fringe frequency is linearly dependent on the glare point spacing (see Equation \ref{Equ:IPI_formula_general}), a smaller glare point spacing results in further spread apart fringes and hence, less fringes within the particle image, compare Figure \ref{fig:IPI_WorkingPrinciple} (analogy to a double slit). As a particle grows the glare point spread apart physically, resulting in the imaged fringes to move closer together. The Nyquist frequency, consequently, restricts the maximum distance of two glare points (and hence particle size), as the fringe spacing becomes to small to distinguish. As a consequence, the upper limit for the measurable particle size can be increased by choosing scattering angles closer to 180° (as there the glare point spacing is smaller for the same particle size). 

Since for small bubbles the interference pattern is more spread apart (less fringes visible - shorter signal sampled), the uncertainty of the fringe frequency is higher (Fourier trade-off between sampled length of a signal and the frequency uncertainty). Consequently a longer sampled signal (more visible fringes) are desirable for lower uncertainty of the frequency, and therefore also the bubble size. More visible fringes for the same bubble size can be obtained by choosing scattering angles closer to 116° (or 99°) as the glare point separation on the bubble grows and as a consequence the fringes shift closer together. 

\FloatBarrier

\subsection*{Geometrical Optics Approach to the Glare Point position}
\label{Sec:Visibility_Sub_GO}

\begin{figure}  
\centering\subfigure{\includegraphics[width=0.4\textwidth]{Figures/Legend2.pdf}}\\
    \addtocounter{subfigure}{-1}
    \subfigure[water droplet $m=1.333$]{\includegraphics[trim={0 0 0 30},clip,width=0.4\textwidth]{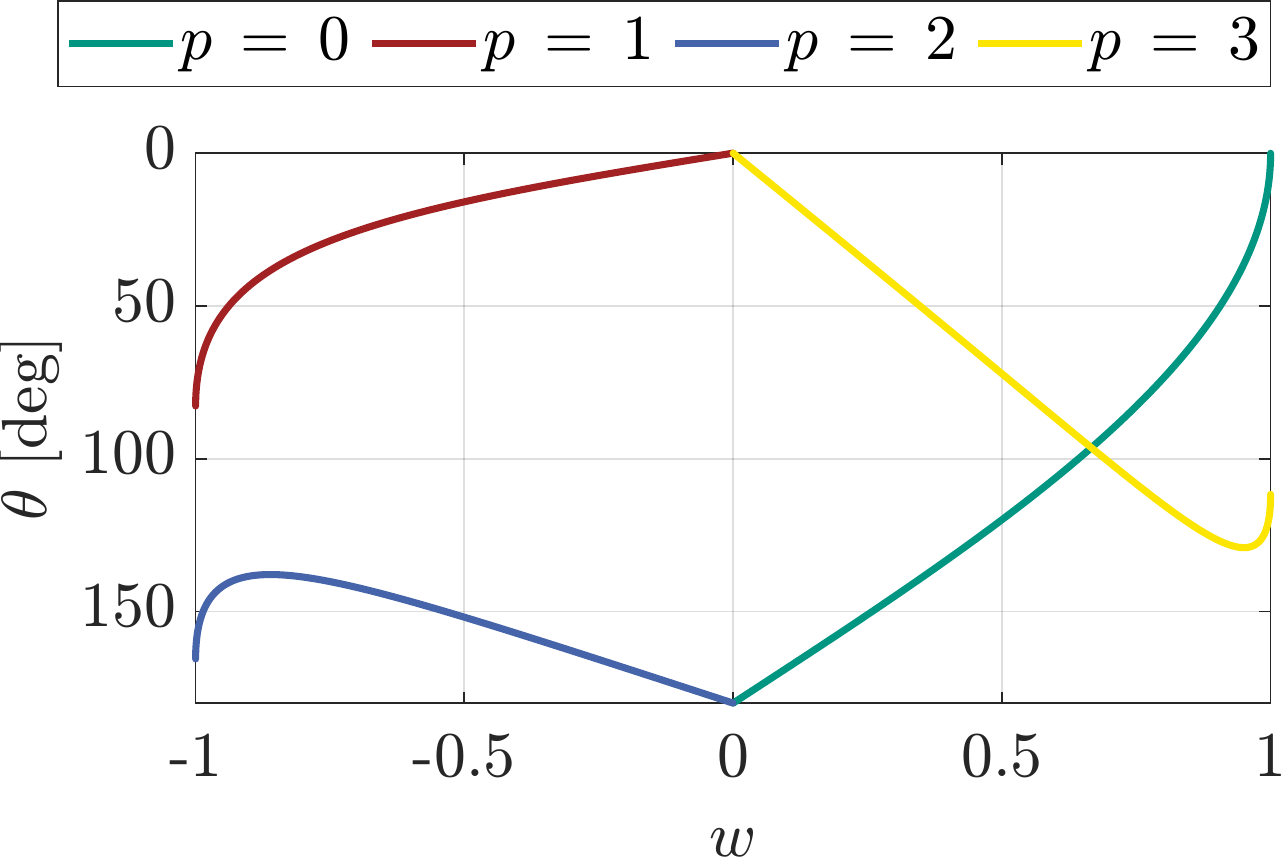}}
    \subfigure[air bubble in water $m=1/1.333$]{\includegraphics[trim={0 0 0 30},clip,width=0.4\textwidth]{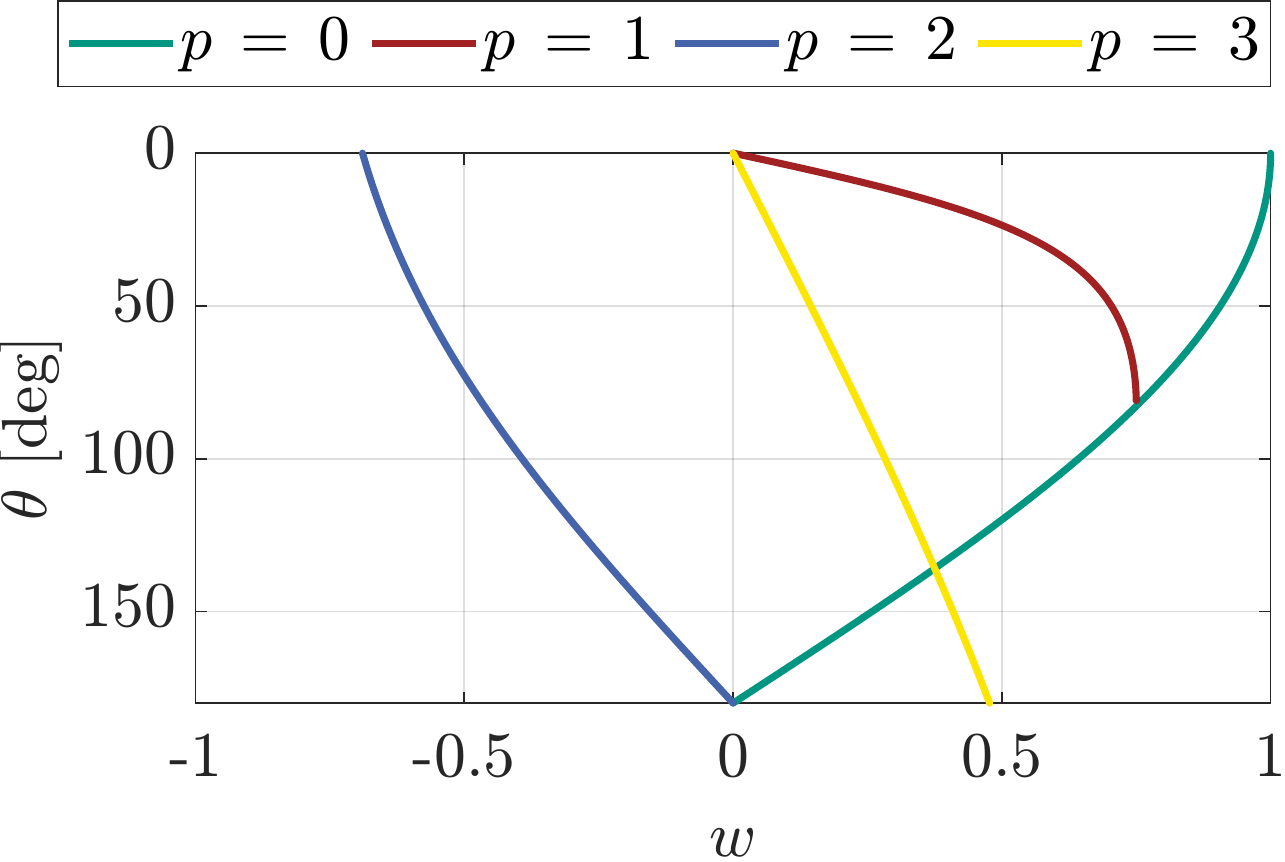}}
    \caption{Glare point position $w$ over the scattering angle visualized for the reflected and first three orders refracted light rays. The glare point map was calculated from GO approximation. The rainbow angles can be visualized in this representation by $d\theta/dw=0$ \cite{vandeHulst.1991}.}
    \label{fig:GP_Maps}
\end{figure}

For the identification of applicable scattering angles the calculation of $A_{\,(w)}$ by means of Debye series expansion was used, since it provided better physical insight compare to the GO approximation. While the glare point position can be derived from these calculations, it is computationally expensive. For an application in which the desired scattering angle is already known, the GO approximation provides a significantly less complex approach to determine the glare point position. Therefore, in this Section a method for the calculation of the glare point position by means of GO will be discussed briefly.\\
Recalling Equation \ref{Equ:TotalDeflec_p1} and using Snell's law \ref{Equ:Snell}, the angles $\beta^{(p)}_i$ and $\beta^{(p)}_t$ can be calculated from the total reflection $\theta'$ in a generalized fashion for any scattering order $p$ \cite{vandeHulst.1991}:
\begin{equation}
    \theta'=2(\beta^{(p)}_i -p\beta^{(p)}_t)=2\pi k + q\theta
    \label{Equ: GO_VDH}
\end{equation}
with $k$ being an integer and $q\pm1$. Since $k$ and $q$ are unknown for the solution, the most forward way to solve the equation, is the calculation of $\theta$ by inserting values for $k$ and $q$, and $\beta^{(p)}_t$. Then the combination of $k$ and $q$ which will return $\theta$ values in the range of 0 to $\pi$ with $\Im\{\theta\}=0$ solve the equation for the order $p$. Values for droplets and bubbles for $p=0,1,2,3$ are provided in Tabular \ref{tab: k_and_q_vaules}. Using Equation \ref{Equ: GO_VDH} and $w=q\cos(\beta^{(p)}_i)$ the glare point position $w$ of order $p$ and be calculated \cite{vandeHulst.1991}, which is shown in a glare point map, see Figure \ref{fig:GP_Maps}. In this glare point map both the glare point separation $\Delta^{(i,j)}_{GP}$ and the position of rainbow angles ($d\theta/dw=0$), which importance has been previously discussed, can be directly determined. Concluding this section, the necessary tools for IPI measurements in front-, side- and back-scatter have been discussed for the calculation of the previously unknown $\Delta^{(i,j)}_{\mathrm{GP}\,(\theta)}$ (glare point paring and spacing).

\begin{center}
\begin{table}[htb]
\centering
\caption{$k$ and $q$ values for the calculation of $\beta^{(p)}_i$ and $\beta^{(p)}_t$}
\begin{tabular}{lccccc}
\toprule
$p$ & Droplet $k$ & Droplet $q$ & Bubble $k$ & Bubble $q$ \\
\midrule
0 & 0 & 1 & 0 & 1\\
1 & 0 & -1 & 0 & 1\\
2 & 0 & -1 & 0 & -1\\
3 & -1 & 1 & 1 & 1\\\bottomrule
\end{tabular}
\label{tab: k_and_q_vaules}
\end{table}
\end{center}

\FloatBarrier

\section*{Application and experimental Validation}
\label{Sec:Experiment}

In this section, the previously derived methods will be applied to real IPI experiments on air bubbles in water under different scattering angles in the front-, side- and back-scatter regime, to proof validity of the theoretical approach. Furthermore it functions as an example for application of the methods.\\
The experimental setup comprises of a rectangular water tank made from glass ($n_glas=1.52$, size 20x20x20\,cm) which was filled with distilled water to avoid the presence of other particles besides bubbles. Salt was added to the distilled water to obtain a electrically conductive solution. An anode and cathode were added to generate Hydrogen bubbles ($n_1=1$) by means of electrolysis. The anode, on which the bubbles were generated, is a steel rod ending in an thin hook shaped (u-shape) wire. The hook shape allowed the anode to be placed from above inside the water while maintaining open space for the bubbles to detach from the wire emerging into the field of view. The anode was sheathed in isolating black rubber, except for the last 0.5\,mm of the wire (hook) tip, to ensure only bubble generation at the tip of the hook and avoid reflections from the metal. The cathode was placed in one of the corners of the water tank outside the field of view (also covered in black isolating material to avoid reflection except of the tip). Applying constant current between the anode and cathode allowed for a constant generation of bubbles with a fixed size distribution. Consequently, measuring the bubble-size probability density function (pdf) from different scattering angles, will result in overlapping pdfs, for a correct theory. 

\begin{figure}
    \centering
    \subfigure[schematic sketch]{\includegraphics[width=0.4\textwidth]{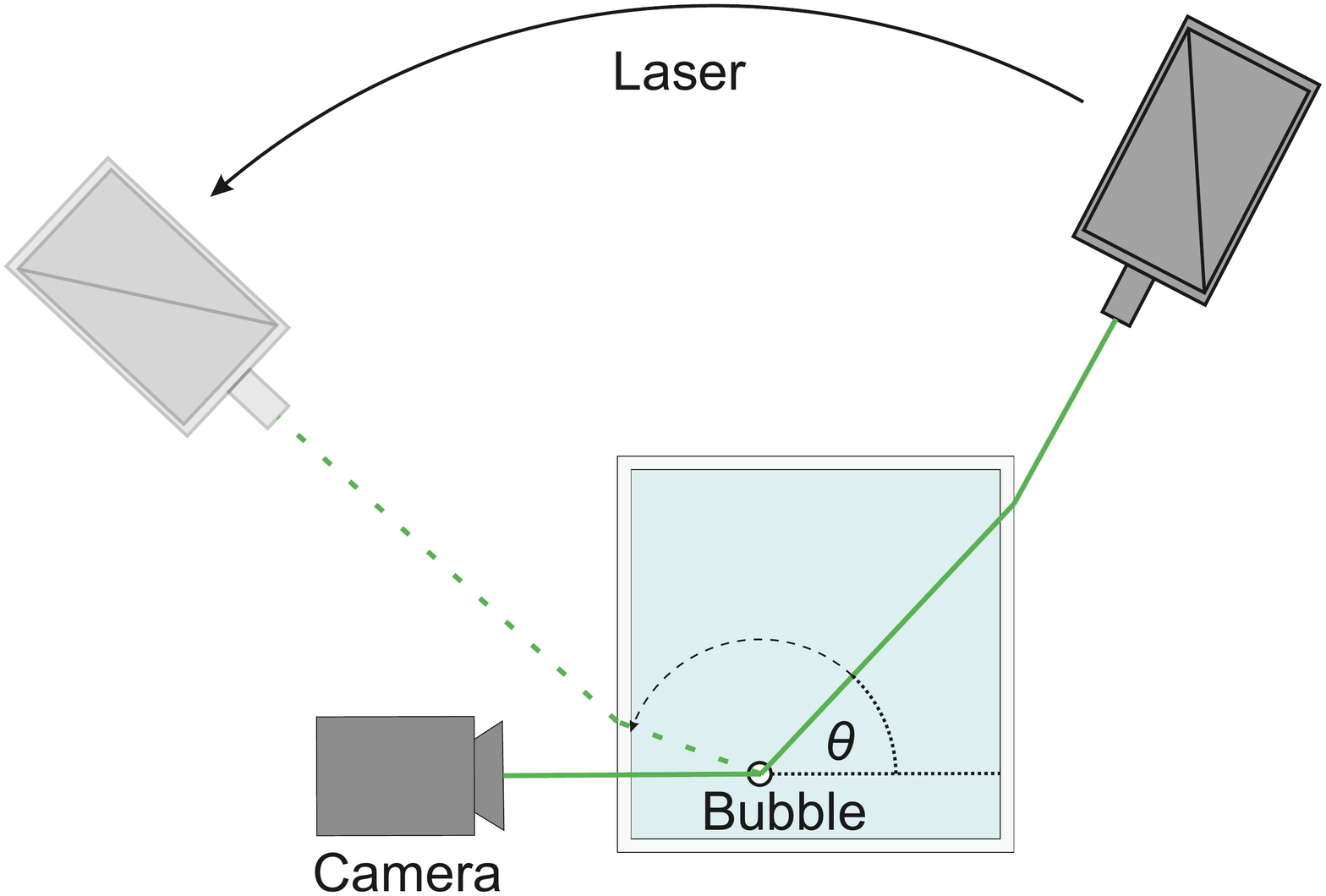}}
    \subfigure[experimental setup]{\includegraphics[width=0.4\textwidth]{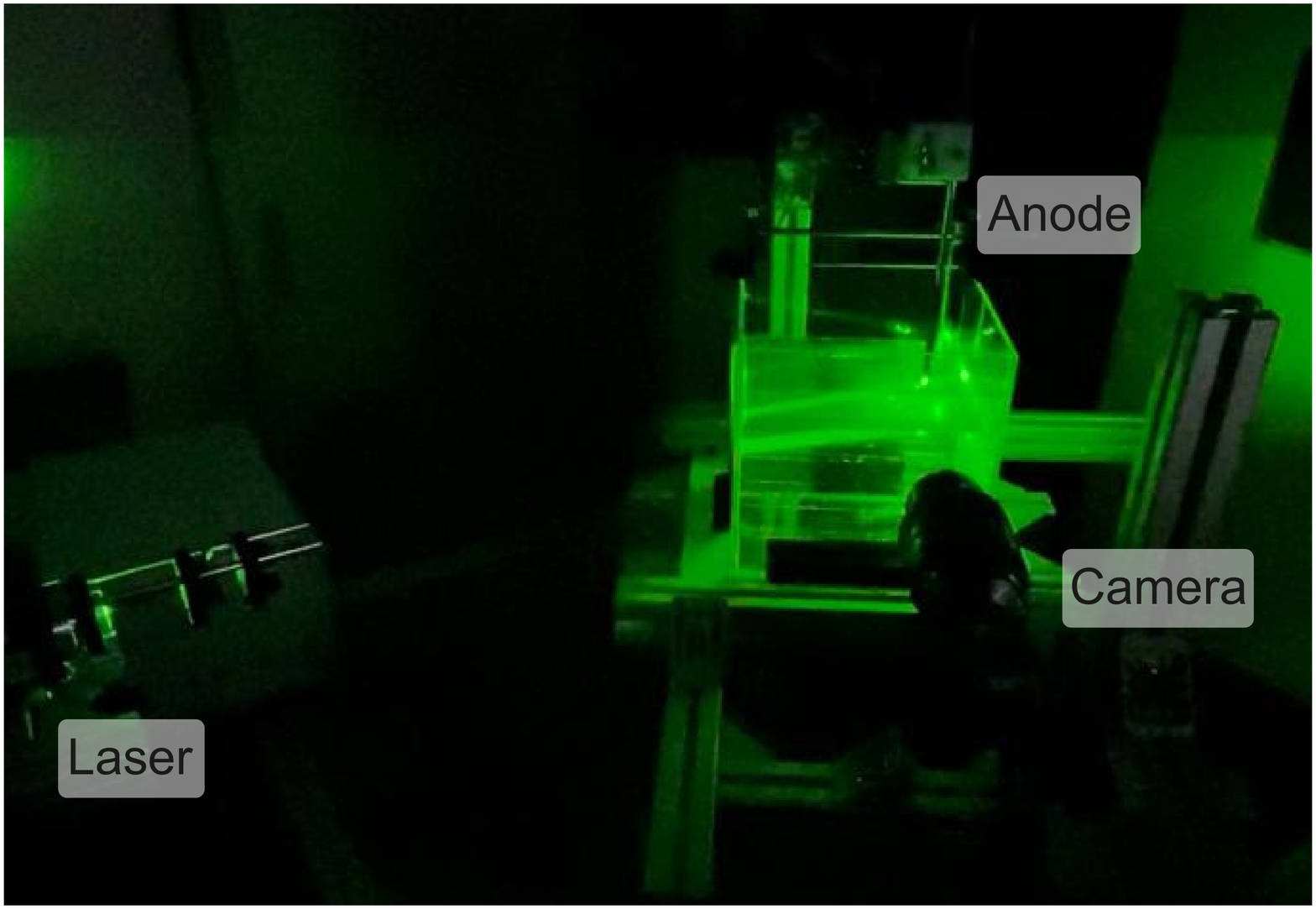}}
    \caption{Validation experiment a) schematic principle b) real experimental setup: The camera images bubbles in a rectangular water tank generated by a point source (electrolysis). The laser illuminating the bubble could be rotated relative to the camera and water tank and consequently, the scattering angle could be adjusted to a 180° degree range.}
    \label{fig:ValidationExp_Setup}
\end{figure}

The whole setup was build in order for the laser illuminating the bubbles to be rotated relative to the tank and the camera. The camera was fixed with the water tank, facing one of the glass walls at an 90° angle, see Figure \ref{fig:ValidationExp_Setup}. By rotation of the setup, the scattering angle was be varied in the whole $\theta=$0..180° range. 
The bubbles were illuminated with an Quantel Evergreen Nd:YAG laser ($\lambda$ = 532\,nm, 200\,mJ/pulse) and images with an PCO Pixelfly camera (CCD-chip, 1392x1040\,pixel, pixel size 6.45x6.45\,$\mu m^2$ ), in an single-image single-pulse setup. The camera was equipped with a Nikon Micro-Nikkor imaging lens (focal length 105\,mm, chosen aperture number APN=4). The whole optical system of the light traveling from the bubble through the water, glass, imaging lens onto the camera chip in an out of focus position is described by the total ray transfer matrix ($B_{tot}=0.02599\,$m). For each scattering angle 1200 images (14bit tiff, 1392x1040\,pixel) were recorded resulting in the order of 12.000-20.000 recorded bubbles per angle.\\

\begin{figure}
    \centering
    \subfigure
    {\includegraphics[trim={0 1cm 0 0},clip,width=0.25\textwidth]{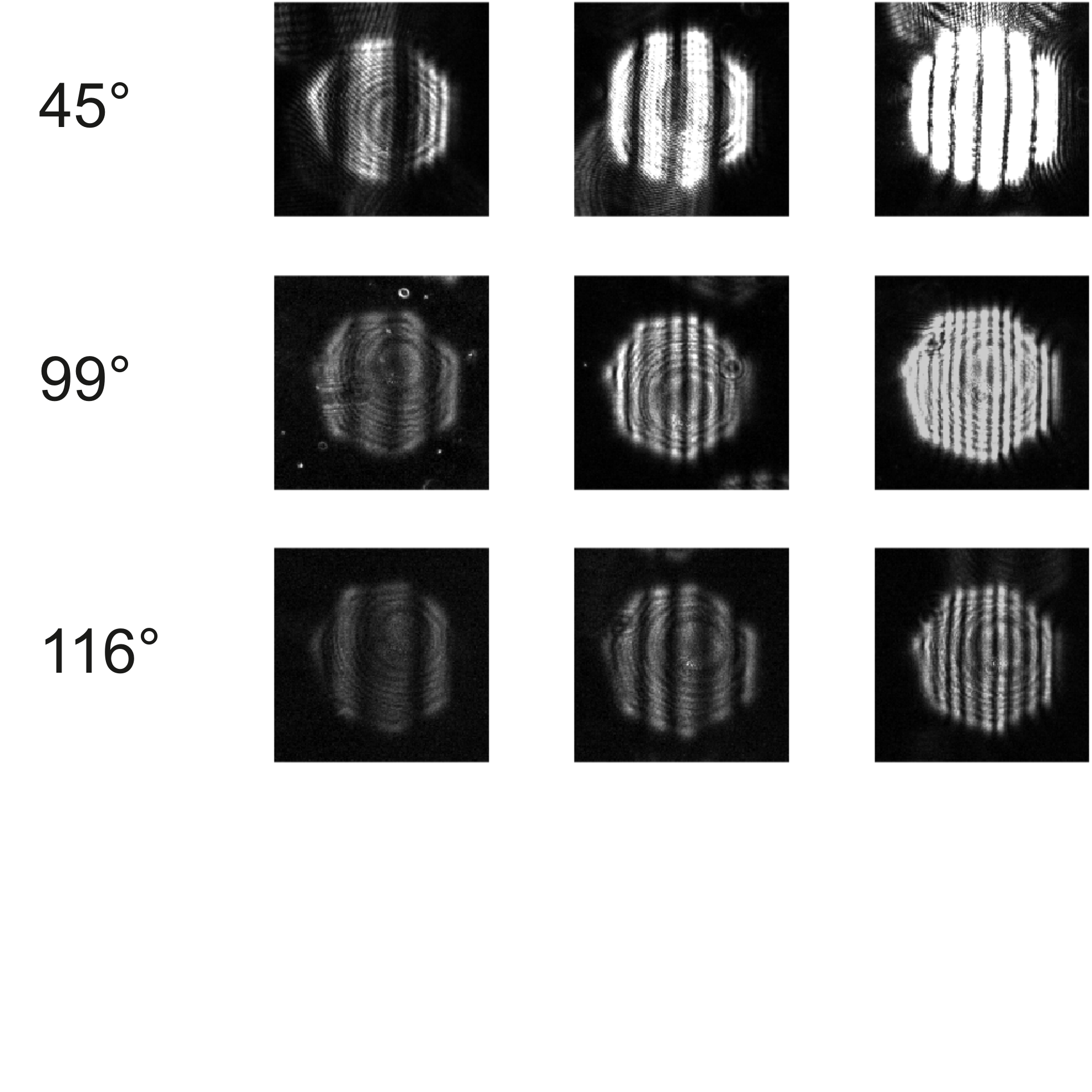}}
    \hspace{1cm}
    \subfigure{\includegraphics[width=0.3\textwidth]{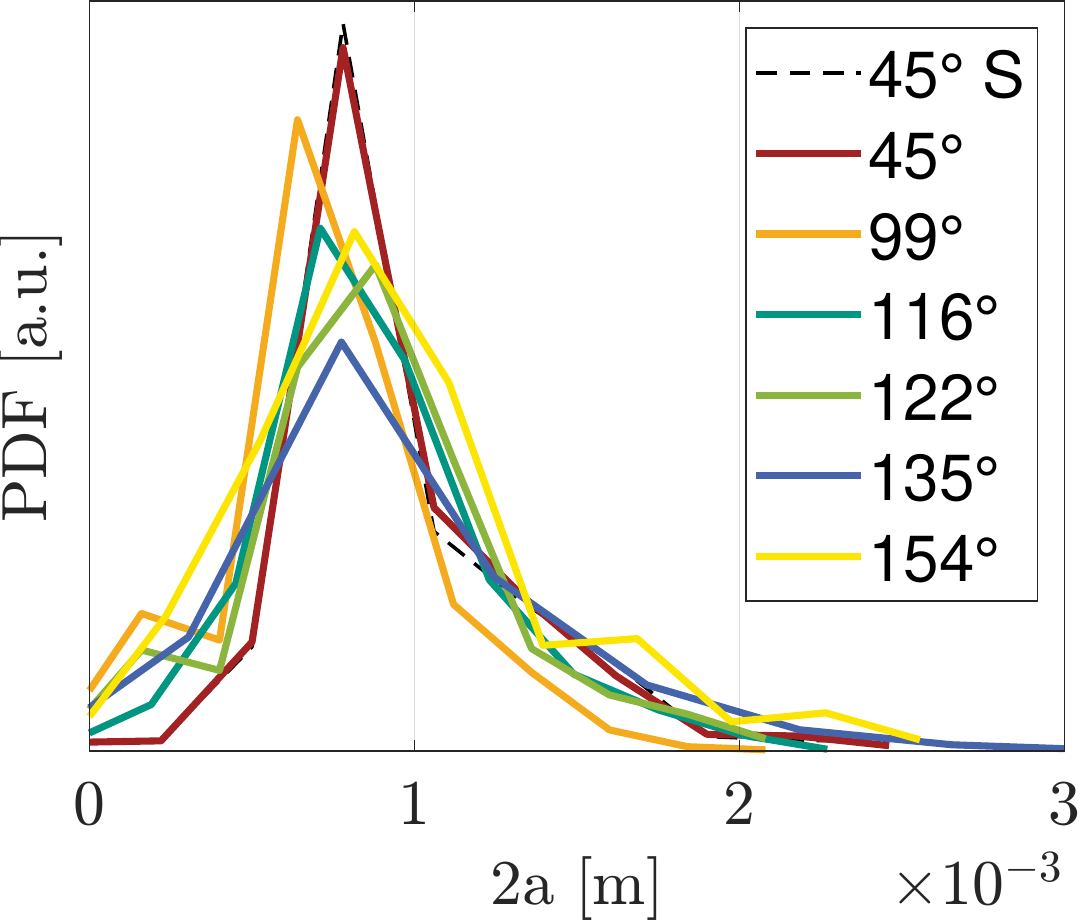}}
    \hspace{1cm}
    \subfigure{\includegraphics[trim={0 1cm 0 0},clip,width=0.25\textwidth]{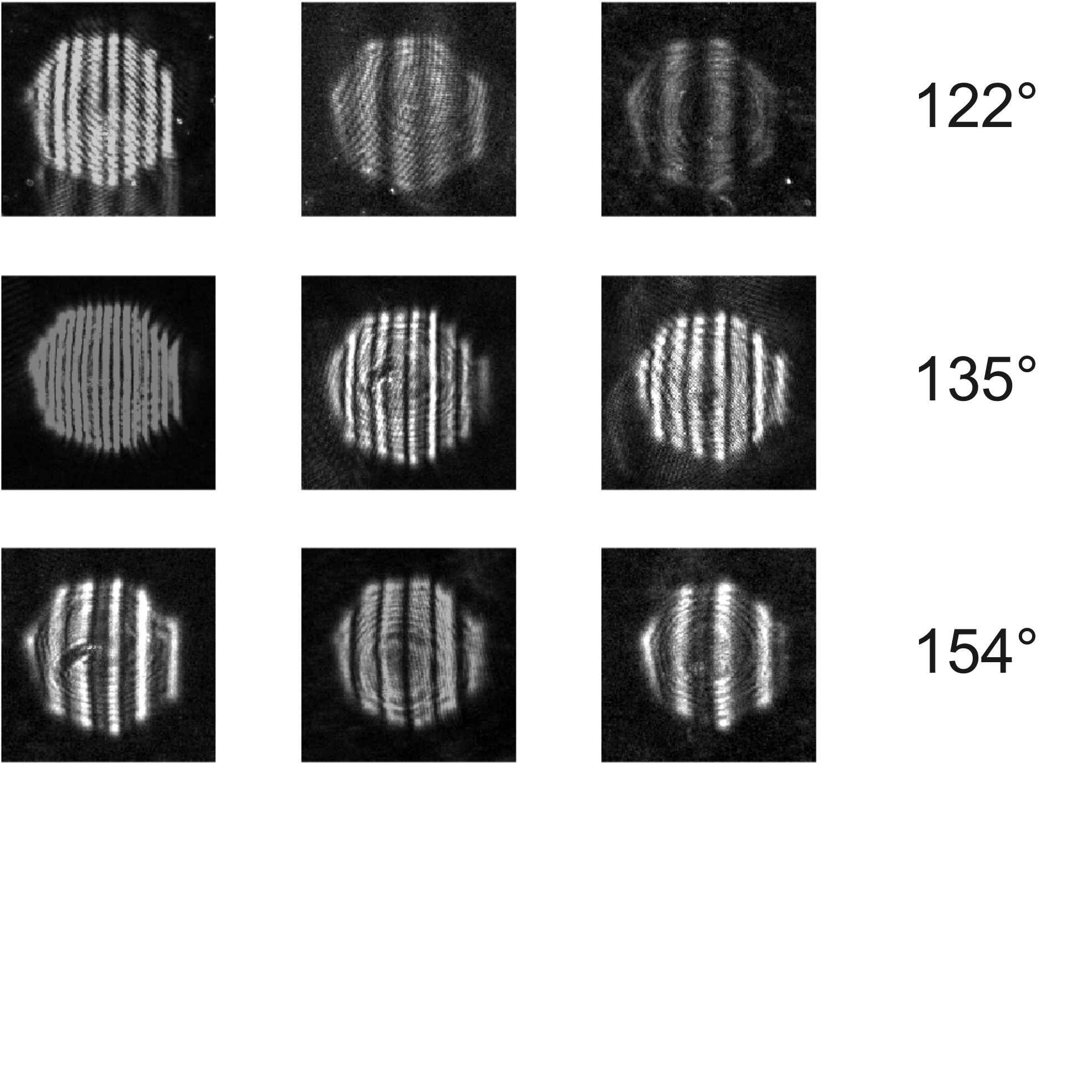}}
    \caption{Results of the IPI bubble sizing experiment for different scattering angles. Depicted is the probability density function (PDF) of the bubble size distribution of the bubble generator for the scattering angles $\theta=$45° (for the calculation by the number of stripes formula ($S$), see Equation \ref{Equ:IPI_classic_bubble} and for the calculation by the frequency formulation, see Equation \ref{Equ:IPI_formula_general}), 99°, 116°, 122°, 135° and 154°. For each angle three exemplary fringe patterns of particles are shown.}
    \label{fig:BubbelPDFs}
\end{figure}

Similar to Section \ref{Sec:Visibility_Sub:GP_Bubble} scattering angles of 45°, 99°, 116°, 122°, 135° and 154° were chosen for measuring the bubble size distribution. The images were pre-processed by means of min-image subtraction and a Laplacian filter to enhance the edge features of the fringes without changing the fringe position for the fringe frequency extraction. The fringe frequency extraction was conducted by a fast Fourier transform (FFT), with low pass filtering in the frequency domain to cut out frequencies caused by noise. Then the base mode in the frequency domain was selected to ignore superimposed frequencies by parasite glare points. Since the FFT lacks adequate frequency resolution for very low frequencies, starting the convergence from the base mode, a first order Fourier function was fitted (first order to ensure only one frequency -the base mode- is fitted), to extract the frequency with higher accuracy. For redundancy, for $\theta=$45° both the number of stripes formulation (Equation \ref{Equ:IPI_classic_bubble}, marked with $S$) as well as the more general fringe frequency formulation (Equation \ref{Equ:IPI_formula_general} was used for the bubble size determination. The bubble size distribution for the different angles are shown in Figure \ref{fig:BubbelPDFs}. It can be seen that the for 45° the distribution of determination by Equation \ref{Equ:IPI_classic_bubble} and Equation \ref{Equ:IPI_formula_general} give the same results, proving their interchangeability. For higher angles only Equation \ref{Equ:IPI_formula_general} is applicable due to the limitations of Equation \ref{Equ:IPI_classic_bubble}. Furthermore for all scattering angles, which were predicted to result in good visibility, clear interference patterns were visible, proving the theory from the visibility considerations (see Equation \ref{Equ: Visibility general}). It shows in Figure \ref{fig:BubbelPDFs} that for all angles similar bubble size distributions were measured, proving the method presented in this work to be working. 

\FloatBarrier

\section*{Concluding Remarks}
\label{Sec:Conclusion}

Interferometric particle imaging is a powerful technique for the sizing of multi-dispersed particles, which can be combined with simultaneous tracking. Due to the conventional method being limited to the forward- and side-scatter regime, the method was limited to applications allowing for an second optical access. In the present work a methodology for the identification of scattering angles in all scattering regimes was presented, both for droplets and bubbles. For the identification of applicable scattering angles, first the visibility of an interference pattern for a glare point paring is investigated. This can be done by visibility plots derived from the scattering intensity calculated by the Debye series expansion of the Mie theory. The visibility plot provides information on the visibility of interference patterns at all scattering angles. For each angle of interest then the glare points should be analyzed in more detail. This can be done be determining the glare point intensity from the complex amplitude from the the Debye calculation. Especially the different suitability of a scattering angles for IPI depending on the light polarization (parallel vs. perpendicular) need to be considered. A less complex and faster, however in some cases over simplified alternative is the investigation with glare point maps which can be derived directly from geometrical optics. The ramification of a single scattering order into multiple glare points needs to be considered beyond the visibility plot and can be avoided at the rainbow angles. After finding an applicable scattering angle and glare point paring, the glare point separation can be determined. The glare point spacing then allows for the derivation of the particle size from the fringe pattern at the given scattering angle. By changing the scattering angle, the glare point spacing can be modified, changing the upper (Nyquist frequency) and lower (frequency uncertainty) limits on particle sizes to be measured with IPI.\\
With this frame work IPI can be measured in all scattering regimes, yielding the method more flexible and allowing for IPI requiring only a single optical access. This will hopefully promote the use of IPI in a broader range of applications and research fields. 

\FloatBarrier

\section*{Competing interests}

The authors declare no competing interests. 

\section*{Data Availability Statement}

All research data related to the present work will be uploaded at KITOpen upon publication. 


\bibliography{main}

\begin{thebibliography}{10}
\urlstyle{rm}
\expandafter\ifx\csname url\endcsname\relax
  \def\url#1{\texttt{#1}}\fi
\expandafter\ifx\csname urlprefix\endcsname\relax\def\urlprefix{URL }\fi
\expandafter\ifx\csname doiprefix\endcsname\relax\def\doiprefix{DOI: }\fi
\providecommand{\bibinfo}[2]{#2}
\providecommand{\eprint}[2][]{\url{#2}}

\bibitem{Maeda2002}
\bibinfo{author}{Maeda, M.}, \bibinfo{author}{Akasaka, Y.} \&
  \bibinfo{author}{Kawaguchi, T.}
\newblock \bibinfo{journal}{\bibinfo{title}{Improvements of the interferometric
  technique for simultaneous measurement of droplet size and velocity vector
  field and its application to a transient spray}}.
\newblock {\emph{\JournalTitle{Experiments in Fluids}}}
  \textbf{\bibinfo{volume}{33}}, \bibinfo{pages}{125--134},
  \doiprefix\url{https://doi.org/10.1007/s00348-002-0453-4}
  (\bibinfo{year}{2002}).

\bibitem{Hardalupas2009}
\bibinfo{author}{Hardalupas, Y.}, \bibinfo{author}{Sahu, S.},
  \bibinfo{author}{Taylor, A.} \& \bibinfo{author}{Zarogoulidis, K.}
\newblock \bibinfo{journal}{\bibinfo{title}{Simultaneous planar measurement of
  droplet velocity and size with gas phase velocities in a spray by combined
  ilids and piv techniques}}.
\newblock {\emph{\JournalTitle{Experiments in Fluids}}}
  \textbf{\bibinfo{volume}{49}}, \bibinfo{pages}{417--434},
  \doiprefix\url{https://doi.org/10.1007/s00348-009-0802-7}
  (\bibinfo{year}{2010}).

\bibitem{Matsuura2006}
\bibinfo{author}{Matsuura, K.} \emph{et~al.}
\newblock \bibinfo{title}{Simultaneous planar measurement of size and
  three-component velocity of droplets in an aero-engine airblast fuel spray by
  stereoscopic interferometric laser imaging technique} (\bibinfo{year}{2006}).

\bibitem{Fujisawa2003}
\bibinfo{author}{Fujisawa, N.}, \bibinfo{author}{Hosokawa, A.} \&
  \bibinfo{author}{Tomimatsu, S.}
\newblock \bibinfo{journal}{\bibinfo{title}{Simultaneous measurement of droplet
  size and velocity field by an interferometric imaging technique in spray
  combustion}}.
\newblock {\emph{\JournalTitle{Measurement Science and Technology}}}
  \textbf{\bibinfo{volume}{14}}, \bibinfo{pages}{1341--1349},
  \doiprefix\url{https://doi.org/10.1088/0957-0233/14/8/320}
  (\bibinfo{year}{2003}).

\bibitem{Dunker2016}
\bibinfo{author}{Dunker, C.}, \bibinfo{author}{Roloff, C.} \&
  \bibinfo{author}{Grassmann, A.}
\newblock \bibinfo{journal}{\bibinfo{title}{Interferometric laser imaging for
  in-flight cloud droplet sizing}}.
\newblock {\emph{\JournalTitle{Measurement Science and Technology}}}
  \textbf{\bibinfo{volume}{27}},
  \doiprefix\url{https://doi.org/10.1088/0957-0233/27/12/124004}
  (\bibinfo{year}{2016}).

\bibitem{KIELAR2016}
\bibinfo{author}{{Jacquot Kielar}, J.} \emph{et~al.}
\newblock \bibinfo{journal}{\bibinfo{title}{Size determination of mixed liquid
  and frozen water droplets using interferometric out-of-focus imaging}}.
\newblock {\emph{\JournalTitle{Journal of Quantitative Spectroscopy and
  Radiative Transfer}}} \textbf{\bibinfo{volume}{178}},
  \bibinfo{pages}{108--116},
  \doiprefix\url{https://doi.org/10.1016/j.jqsrt.2015.09.009}
  (\bibinfo{year}{2016}).
\newblock \bibinfo{note}{Electromagnetic and light scattering by nonspherical
  particles XV: Celebrating 150 years of Maxwell's electromagnetics}.

\bibitem{Querel}
\bibinfo{author}{Quérel, A.}, \bibinfo{author}{Lemaitre, P.},
  \bibinfo{author}{Brunel, M.}, \bibinfo{author}{Porcheron, E.} \&
  \bibinfo{author}{Gréhan, G.}
\newblock \bibinfo{journal}{\bibinfo{title}{Real-time global interferometric
  laser imaging for the droplet sizing (ilids) algorithm for airborne
  research}}.
\newblock {\emph{\JournalTitle{Measurement Science and Technology}}}
  \textbf{\bibinfo{volume}{21}}, \bibinfo{pages}{015306},
  \doiprefix\url{https://doi.org/10.1088/0957-0233/21/1/015306}
  (\bibinfo{year}{2009}).

\bibitem{Lacagnina2011}
\bibinfo{author}{Lacagnina, G.}, \bibinfo{author}{Grizzi, S.},
  \bibinfo{author}{Falchi, M.}, \bibinfo{author}{Felice, F.~D.} \&
  \bibinfo{author}{Romano, G.~P.}
\newblock \bibinfo{journal}{\bibinfo{title}{Simultaneous size and velocity
  measurements of cavitating microbubbles using interferometric laser
  imaging}}.
\newblock {\emph{\JournalTitle{Experiments in Fluids}}}
  \textbf{\bibinfo{volume}{50}}, \bibinfo{pages}{1153--1167},
  \doiprefix\url{https://doi.org/10.1007/S00348-011-1055-9}
  (\bibinfo{year}{2011}).

\bibitem{Kawaguchi.2002}
\bibinfo{author}{Kawaguchi, T.}, \bibinfo{author}{Akasaka, Y.} \&
  \bibinfo{author}{Maeda, M.}
\newblock \bibinfo{journal}{\bibinfo{title}{Size measurements of droplets and
  bubbles by advanced interferometric laser imaging technique}}.
\newblock {\emph{\JournalTitle{Measurement Science and Technology}}}
  \textbf{\bibinfo{volume}{13}}, \bibinfo{pages}{308},
  \doiprefix\url{https://doi.org/10.1088/0957-0233/13/3/312}
  (\bibinfo{year}{2002}).

\bibitem{Niwa.2000}
\bibinfo{author}{Niwa, Y.}, \bibinfo{author}{Kamiya, Y.},
  \bibinfo{author}{Kawaguchi, T.} \& \bibinfo{author}{Maeda, M.}
\newblock \bibinfo{journal}{\bibinfo{title}{Bubble sizing by interferometric
  laser imaging}}.
\newblock {\emph{\JournalTitle{10th International Symposium on Application of
  Laser Techniques to Fluid Mechanics}}}  (\bibinfo{year}{2000}).

\bibitem{Konig.1986}
\bibinfo{author}{K{\"o}nig, G.}, \bibinfo{author}{Anders, K.} \&
  \bibinfo{author}{Frohn, A.}
\newblock \bibinfo{journal}{\bibinfo{title}{A new light-scattering technique to
  measure the diameter of periodically generated moving droplets}}.
\newblock {\emph{\JournalTitle{Journal of Aerosol Science}}}
  \textbf{\bibinfo{volume}{17}}, \bibinfo{pages}{157--167},
  \doiprefix\url{https://doi.org/10.1016/0021-8502(86)90063-7}
  (\bibinfo{year}{1986}).

\bibitem{Glover.95}
\bibinfo{author}{Glover, A.~R.}, \bibinfo{author}{Skippon, S.~M.} \&
  \bibinfo{author}{Boyle, R.~D.}
\newblock \bibinfo{journal}{\bibinfo{title}{Interferometric laser imaging for
  droplet sizing: a method for droplet-size measurement in sparse spray
  systems}}.
\newblock {\emph{\JournalTitle{Appl. Opt.}}} \textbf{\bibinfo{volume}{34}},
  \bibinfo{pages}{8409--8421},
  \doiprefix\url{https://doi.org/10.1364/AO.34.008409} (\bibinfo{year}{1995}).

\bibitem{Willert.1992}
\bibinfo{author}{Willert, C.~E.} \& \bibinfo{author}{Gharib, M.}
\newblock \bibinfo{journal}{\bibinfo{title}{Three-dimensional particle imaging
  with a single camera}}.
\newblock {\emph{\JournalTitle{Experiments in Fluids}}}
  \textbf{\bibinfo{volume}{12}}, \bibinfo{pages}{353--358}
  (\bibinfo{year}{1992}).

\bibitem{Fuchs.2016}
\bibinfo{author}{Fuchs, T.}, \bibinfo{author}{Hain, R.} \&
  \bibinfo{author}{K{\"a}hler, C.~J.}
\newblock \bibinfo{journal}{\bibinfo{title}{In situ calibrated defocusing ptv
  for wall-bounded measurement volumes}}.
\newblock {\emph{\JournalTitle{Measurement Science and Technology}}}
  \textbf{\bibinfo{volume}{27}}, \bibinfo{pages}{084005}
  (\bibinfo{year}{2016}).

\bibitem{Tropea2007}
\bibinfo{author}{Tropea, C.}, \bibinfo{author}{Yarin, A.} \&
  \bibinfo{author}{Foss, J.}
\newblock \emph{\bibinfo{title}{Springer Handbook of Experimental Fluid
  Mechanics}} (\bibinfo{publisher}{Springer}, \bibinfo{year}{2007}).

\bibitem{Rousselle1999}
\bibinfo{author}{Mouna{\"i}m-Rousselle, C.} \& \bibinfo{author}{Pajot, O.}
\newblock \bibinfo{journal}{\bibinfo{title}{Droplet sizing by mie scattering
  interferometry in a spark ignition engine}}.
\newblock {\emph{\JournalTitle{Particle \& Particle Systems Characterization}}}
  \textbf{\bibinfo{volume}{16}}, \bibinfo{pages}{160--168},
  \doiprefix\url{https://doi.org/10.1002/(SICI)1521-4117(199908)16:4<160::AID-PPSC160>3.0.CO;2-G}
  (\bibinfo{year}{1999}).

\bibitem{Zhang.2018}
\bibinfo{author}{Zhang, H.}, \bibinfo{author}{Wang, X.}, \bibinfo{author}{Sun,
  J.}, \bibinfo{author}{Jia, D.} \& \bibinfo{author}{Liu, T.}
\newblock \bibinfo{journal}{\bibinfo{title}{Multidispersed bubble-size
  measurements by interferometric particle imaging at scattering angles of
  90\&\#x00b0; and 45\&\#x00b0;}}.
\newblock {\emph{\JournalTitle{Appl. Opt.}}} \textbf{\bibinfo{volume}{57}},
  \bibinfo{pages}{10496--10504},
  \doiprefix\url{https://doi.org/10.1364/AO.57.010496} (\bibinfo{year}{2018}).

\bibitem{Russell2020}
\bibinfo{author}{Russell, P.}, \bibinfo{author}{Venning, J.~A.},
  \bibinfo{author}{Pearce, B.~W.} \& \bibinfo{author}{Brandner, P.~A.}
\newblock \bibinfo{journal}{\bibinfo{title}{Calibration of mie scattering
  imaging for microbubble measurement in hydrodynamic test facilities}}.
\newblock {\emph{\JournalTitle{Experiments in Fluids}}}
  \textbf{\bibinfo{volume}{61}},
  \doiprefix\url{https://doi.org/10.1007/s00348-020-2927-7}
  (\bibinfo{year}{2020}).

\bibitem{Semidetnov2003}
\bibinfo{author}{Semidetnov, N.} \& \bibinfo{author}{Tropea, C.}
\newblock \bibinfo{journal}{\bibinfo{title}{Conversion relationships for
  multidimensional particle sizing techniques}}.
\newblock {\emph{\JournalTitle{Measurement Science and Technology}}}
  \textbf{\bibinfo{volume}{15}}, \bibinfo{pages}{112 -- 118},
  \doiprefix\url{https://doi.org/10.1088/0957-0233/15/1/015}
  (\bibinfo{year}{2003}).

\bibitem{vandeHulst.1991}
\bibinfo{author}{{van de Hulst}, H.~C.} \& \bibinfo{author}{Wang, R.~T.}
\newblock \bibinfo{journal}{\bibinfo{title}{Glare points}}.
\newblock {\emph{\JournalTitle{Applied optics}}} \textbf{\bibinfo{volume}{30}},
  \bibinfo{pages}{4755--4763},
  \doiprefix\url{https://doi.org/10.1364/AO.30.004755} (\bibinfo{year}{1991}).

\bibitem{Roth1994}
\bibinfo{author}{Roth, N.}, \bibinfo{author}{Anders, K. D.~I.} \&
  \bibinfo{author}{Frohn, A.}
\newblock \bibinfo{journal}{\bibinfo{title}{Determination of size, evaporation
  rate and freezing of water droplets using light scattering and radiation
  pressure}}.
\newblock {\emph{\JournalTitle{Particle \& Particle Systems Characterization}}}
  \textbf{\bibinfo{volume}{11}}, \bibinfo{pages}{207--211},
  \doiprefix\url{https://doi.org/10.1002/ppsc.19940110307}
  (\bibinfo{year}{1994}).

\bibitem{Madsen.2003}
\bibinfo{author}{Madsen, J.} \emph{et~al.}
\newblock \bibinfo{journal}{\bibinfo{title}{Measurement of droplet size and
  velocity distributions in sprays using interferometric particle imaging (ipi)
  and particle tracking velocimetry (ptv)}}.
\newblock {\emph{\JournalTitle{Conference: Proceedings 9th International
  Conference on Liquid Atomization and Spray Systems - ICLASS 2003}}}
  (\bibinfo{year}{2003}).

\bibitem{born_wolf_2019}
\bibinfo{author}{Born, M.} \& \bibinfo{author}{Wolf, E.}
\newblock \emph{\bibinfo{title}{Principles of Optics: 60th Anniversary
  Edition}} (\bibinfo{publisher}{Cambridge University Press},
  \bibinfo{year}{2019}).

\bibitem{Shen.2012}
\bibinfo{author}{Shen, H.}, \bibinfo{author}{Co{\"e}tmellec, S.},
  \bibinfo{author}{Grehan, G.} \& \bibinfo{author}{Brunel, M.}
\newblock \bibinfo{journal}{\bibinfo{title}{Interferometric laser imaging for
  droplet sizing revisited: elaboration of transfer matrix models for the
  description of complete systems.}}
\newblock {\emph{\JournalTitle{Applied optics}}} \textbf{\bibinfo{volume}{51
  22}}, \bibinfo{pages}{5357--68},
  \doiprefix\url{https://doi.org/10.1364/AO.51.005357} (\bibinfo{year}{2012}).

\bibitem{vandeHulst1957}
\bibinfo{author}{{van de Hulst}, H.~C.} \& \bibinfo{author}{Twersky, V.}
\newblock \emph{\bibinfo{title}{Light Scattering by Small Particles: volume
  10.}} (\bibinfo{publisher}{Courier Corporation, 2012}, \bibinfo{year}{1957 //
  1958}).

\bibitem{Courant1953}
\bibinfo{author}{Courant, R.} \& \bibinfo{author}{Hilbert, D.}
\newblock \emph{\bibinfo{title}{Methods of Mathematical Physics}}
  (\bibinfo{publisher}{{Volume 1, New York: Interscience Publischer, Inc}},
  \bibinfo{year}{1953}).

\bibitem{Abramowitz1965}
\bibinfo{author}{Abramowitz, M.} \& \bibinfo{author}{Stegun, I.~A.}
\newblock \emph{\bibinfo{title}{Handbook of mathematical funktions}}
  (\bibinfo{publisher}{Dover}, \bibinfo{address}{New York N.Y.},
  \bibinfo{year}{1965}).

\bibitem{Bohren1998b}
\bibinfo{author}{Bohren, C.~F.} \& \bibinfo{author}{Huffman, D.~R.}
\newblock \emph{\bibinfo{title}{Absorption and Scattering of Light by Small
  Particles}} (\bibinfo{publisher}{Wiley}, \bibinfo{year}{1998}).

\bibitem{Hovenac1992}
\bibinfo{author}{Hovenac, E.~A.} \& \bibinfo{author}{Lock, J.~A.}
\newblock \bibinfo{journal}{\bibinfo{title}{Assessing the contributions of
  surface waves and complex rays to far-field mie scattering by use of the
  debye series}}.
\newblock {\emph{\JournalTitle{Journal of the Optical Society of America A}}}
  \textbf{\bibinfo{volume}{9}}, \bibinfo{pages}{781},
  \doiprefix\url{https://doi.org/10.1364/JOSAA.9.000781}
  (\bibinfo{year}{1992}).

\bibitem{Gouesbet2003}
\bibinfo{author}{Gouesbet, G.}
\newblock \bibinfo{journal}{\bibinfo{title}{Debye series formulation for
  generalized lorenz-mie theory with the bromwich method}}.
\newblock {\emph{\JournalTitle{Particle {\&} Particle Systems
  Characterization}}} \textbf{\bibinfo{volume}{20}}, \bibinfo{pages}{382--386},
  \doiprefix\url{https://doi.org/10.1002/ppsc.200300886}
  (\bibinfo{year}{2003}).

\bibitem{Dehaeck2008}
\bibinfo{author}{Dehaeck, S.} \& \bibinfo{author}{van Beeck, J.}
\newblock \bibinfo{journal}{\bibinfo{title}{Multifrequency interferometric
  particle imaging for gas bubble sizing}}.
\newblock {\emph{\JournalTitle{Experiments in Fluids}}}
  \textbf{\bibinfo{volume}{45}}, \bibinfo{pages}{823--831},
  \doiprefix\url{https://doi.org/10.1007/S00348-008-0502-8}
  (\bibinfo{year}{2008}).

\bibitem{Nolting2013}
\bibinfo{author}{Nolting, W.}
\newblock \emph{\bibinfo{title}{Grundkurs Theoretische Physik 3}}
  (\bibinfo{publisher}{Springer Spektrum Berlin, Heidelberg},
  \bibinfo{year}{2013}).

\end{thebibliography}

\end{document}